\documentclass[11pt, letterpaper]{article}

\usepackage{lineno}
\usepackage{graphicx}
\usepackage{amsmath}
\usepackage{url}
\usepackage{flushend}
\usepackage[latin1]{inputenc}
\usepackage{amssymb}
\usepackage{setspace}
\usepackage{epstopdf}
\usepackage{color}
\usepackage[letterpaper,left=1in,right=1in,top=1in,bottom=1in]{geometry}
\usepackage[linesnumbered,ruled,vlined,longend]{algorithm2e}
\usepackage{multirow}
\usepackage{longtable}
\usepackage{rotating}
\usepackage{threeparttable}
\usepackage{colortbl}
\usepackage{amsthm}
\usepackage{enumerate}
\usepackage{bm}
\usepackage{algorithmic}
\usepackage{soul}
\usepackage{pifont}
\usepackage{alltt}
\usepackage{booktabs}
\usepackage{siunitx}
\usepackage{pbox}
\usepackage{amsfonts,amssymb}
\usepackage{tabularx}
\usepackage{amsthm,amsmath,amssymb}
\usepackage{mathrsfs}
\usepackage[hidelinks]{hyperref}
\usepackage{subfigure}
\usepackage[numbers,sort&compress]{natbib}

\title{\Large \bf State Estimation for One-Dimensional Agro-hydrological Processes with Model Mismatch }

\author{\centerline{\normalsize Zhuangyu Liu$^{a,b}$, Jinfeng Liu$^{b,}$\thanks{Corresponding author: J. Liu. Email: jinfeng@ualberta.ca}\ , Shunyi Zhao$^{a}$, Xiaoli Luan$^{a,}$\thanks{Corresponding author: X. Luan. Email: xlluan@jiangnan.edu.cn}\ , Fei Liu$^{a}$}\vspace{5mm}\\
    \centerline{\small $^{a}$ School of Internet of Things Engineering, Jiangnan University, Wuxi\ 214122, China}\\
    \centerline{\small $^{b}$ Department of Chemical \& Materials Engineering, University of Alberta,}\\
    \centerline{\small Edmonton, AB, Canada, T6G 1H9}}
    
\allowdisplaybreaks

\begin{document}

%\doublespacing
%\onehalfspacing
\date{}

\maketitle
\setstretch{1.39}

\begin{abstract}
	The importance of accurate soil moisture data for the development of modern closed-loop irrigation systems cannot be overstated. Due to the diversity of soil, it is difficult to obtain an accurate model for agro-hydrological system. In this study, soil moisture estimation in 1D agro-hydrological systems with model mismatch is the focus. To address the problem of model mismatch, a nonlinear state-space model derived from the Richards equation is utilized, along with additive unknown inputs.	The determination of the number of sensors required is achieved through sensitivity analysis and the orthogonalization projection method. To estimate states and unknown inputs in real-time, a recursive expectation maximization (EM) algorithm derived from the conventional EM algorithm is employed. During the E-step, the extended Kalman filter (EKF) is used to compute states and covariance in the recursive Q-function, while in the M-step, unknown inputs are updated by locally maximizing the recursive Q-function.	The estimation performance is evaluated using comprehensive simulations. Through this method, accurate soil moisture estimation can be obtained, even in the presence of model mismatch.
\end{abstract}


\section{Introduction} \label{sec:introduction}

The challenges posed by water and food shortages are becoming increasingly significant worldwide, fueled by a combination of population growth and climate change. According to the United Nations, agriculture accounts for approximately $70\%$ of available freshwater usage, with irrigation being the primary contributor to this figure. Fischer et al. report that the current global average efficiency of water consumption in irrigation is approximately $50\%$ \cite{fischer2007climate}. Increasing water-use efficiency in irrigation is essential to alleviate water shortages. However, open-loop irrigation, which wastes a significant amount of water, is still commonly used. Closed-loop irrigation systems offer great potential to reduce the impact on water supplies and improve crop health, but they are not yet widely adopted \cite{mao2018soil}. The implementation of a closed-loop irrigation system necessitates accurate agro-hydrological models and real-time soil moisture data for the entire field. However, obtaining such information is often challenging \cite{nahar2019parameter}. One approach is to estimate soil moisture from limited sensor data. However, the reliability of these estimates largely depends on the quality of the agro-hydrological model used. Unfortunately, due to the complex nature of soil parameters and the discretization of partial differential equations, model mismatch is an unavoidable issue in irrigation models. Therefore, the objective is to devise a state estimation method to efficiently overcome model mismatch and provide reliable soil moisture estimates in this research. 

The dynamics of soil water are explained through the use of the Richards equation, where soil properties are related to several parameters in the equation. Soil moisture is a crucial indicator in agro-hydrological systems. Therefore, implementing soil moisture estimation approach is of utmost importance. Some potential solutions to the problem of estimating soil moisture content has been studied \cite{reichle2002extended,erdal2015importance,zhang2017state,lu2011dual}. As an example, the Kalman filter was utilized to predict soil water content in \cite{walker2001one} based on the linearization of a simple soil profile model. In another study \cite{agyeman2021soil}, the extended Kalman filter was employed to estimate soil moisture in a 3D agro-hydrological polar system. In the abovementioned references, the model mismatch is not considered. On the other hand, precise estimation of soil parameters is also crucial \cite{yin2021consensus}. In \cite{bo2020parameter}, a moving horizon estimation method is introduced to simultaneously estimate the parameters and states of agro-hydrological systems. This method enables the estimation of the capillary pressure head and model parameters. However, it should be noted that not all parameters can be accurately estimated through this approach. In cases where certain parameters cannot be estimated, they are assumed to be constant and given an initial guess. It is possible that the initial guess for certain parameters may not be precise, leading to model mismatch. This, in turn, can result in poor estimates of soil moisture that may fail to accurately track the true value or even diverge altogether. Additionally, external factors such as crop coefficient and evaporation rate can be difficult to estimate accurately, as pointed out in \cite{agyeman2021soil}. And due to fluctuations in the surrounding environment, these parameters could exhibit temporal variation. On occasion, direct estimation of such parameters is either impossible or impractical. The estimation of soil moisture is further complicated by the fact that the Richards equation is a partial differential equation (PDE) \cite{van1980closed}. As a result, soil water content is a function of both spatial and temporal variables, particularly under non-steady flow conditions. The finite difference method is a common approach used to solve PDEs by discretizing them into a set of ODEs. However, this approach can result in model errors during the discretization procedure \cite{bo2020parameter}. As discussed earlier, model mismatch is an inevitable issue in the estimation of the irrigation process, making it crucial to address this problem. Another problem related to the state estimation for agro-hydrological systems is the sensor selection. Optimal sensor placement is a significant aspect of soil moisture estimation. In \cite{sahoo2019optimal}, a systematic strategy was developed to address this problem. Furthermore, the issue of sensor placement for an actual field was tackled in \cite{orouskhani2023impact} by considering the degree of observability. In this study, the model mismatch is addressed by treating it as additive unknown inputs to the state equation, and the sensor selection problem for agro-hydrological systems is addressed.

The estimate of unknown inputs has historically been studied widely in the perspective of linear systems. An observer is presented for dynamic linear systems with unknown inputs, along with adequate requirements for the observer's existence \cite{darouach1994full}. In the multiple-model approach, unknown inputs are represented by a random variable of which transition probabilities follow a known Markov chain \cite{orjuela2009simultaneous}. A recursive method known as the augmented state Kalman filter (ASKF) has been proposed in linear systems, the conventional strategy aims to integrate unknown inputs into the state vector \cite{kong2019internal,kong2020filtering}. However, if the unknown inputs are time-varying, it is difficult to construct the state transition matrix for the unknown inputs. The expectation-maximization (EM) algorithm is a method for jointly estimating the states and identifying the unknown parameters in linear systems, it is an iterative optimization algorithm used for point estimation. Recent studies on the estimation of unknown inputs in nonlinear systems have made significant progress. For instance, in \cite{kim2020simultaneous}, a filter was designed to simultaneously estimate the unknown inputs and states of a class of nonlinear stochastic systems where unknown inputs are arbitrary. Additionally, \cite{khan2019simultaneous} introduced an EM algorithm using a particle smoother in nonlinear system identification to address the estimation of unknown inputs.

The EM algorithm is originally proposed by Dempster in \cite{dempster1977maximum} to address the maximum likelihood estimation under the situation of missing data, then it was developed to handle parameter estimation problems in hidden models widely. The EM algorithm is an iterative optimization methodology, which requires all the historical data incorporated in the computation \cite{guo2017augmented,zhang2014expectation,zhao2012algorithm}. The technique must be completely recalculated when additional sensor data are received, which is a drawback in terms of compute complexity \cite{krishnamurthy1993line}. Motivated by the above discussion and the characteristic of the agro-hydrological system, this paper is dedicated to reduce calculation complexity by recursion and update the model error in a real-time manner. The recursive implementation of the EM algorithm has recently garnered tremendous scholarly interest. A recursive EM (REM) method was first presented by Titterington \cite{titterington1984recursive}. Stochastic gradient strategy is the foundation of the algorithm, online modifications are made via the Fisher information matrix (FIM), the drawback is that the inverse of the FIM is required which is not easy to calculate, especially for high dimensional systems. The conditional expectation of likelihood was obtained using a stochastic approximation step by Capp\'e, which avoided the need for inverting the FIM \cite{cappe2009line,cappe2011online}. In the initial, REM is developed for parameter estimation of a hidden Markov model \cite{ryden1997recursive}. In recent years, several studies have employed the REM method to solve the problem of parameter identification. For nonlinear state-space systems with a jump Markov model, a REM identification strategy was presented in \cite{ozkan2014recursive}. Identification of the hidden Markov models using REM algorithm is concerned, and convergence result of the algorithm was derived in \cite{krishnamurthy2002recursive}.

In this work, we address the problem of state estimation for 1D agro-hydrological system with model mismatch. The model mismatch is treated as the additive unknown inputs to the state equations. Due to the discretization for the whole filed, dozens of states are common for the 1D agro-hydrological process. It is not possible to place sensor for each. To determine the minimum number of sensors and their optimal placement, this study employs the sensitivity analysis and orthogonalization method. After that, we propose a systematic structure for state and unknown inputs estimation by derivating a recursive EM algorithm. In E-step of the REM algorithm, a recursive approximation of the Q-function is represented, and the EKF is employed to update the states and the covariance. The M-step of the proposed algorithm involves the correction of the unknown inputs at each time instance by maximizing the modified recursive Q-function. The main contributions of this work can be summarized as follows:
\begin{itemize}

\item  The model mismatch is inevitable for agro-hydrological system because the soil property parameters are always difficult to determine, and the external environment will affect the system significantly. The problem of state estimation for 1D agro-hydrological system with model mismatch is addressed in this work. 
	
\item It is quite costly to place sensors at all the nodes of an irrigation field. Consequently, the sensor placement problem has tremendous significance for state estimation of agro-hydrological system. The optimal sensor placement and minimum number of sensors are determined in this paper by adopting the sensitivity analysis and orthogonalization approach.
	
\item A recursive EM algorithm is derived from the conventional EM algorithm to implement estimation for states and model mismatch. The recursive EM method has the benefits of real-time and computational efficiency. For the agro-hydrological system or some other high dimensional systems, the recursive EM algorithm is more practical.

\end{itemize}
 
\section{Agro-hydrological System and Problem Formulation}
\label{Section 2}

\subsection{System model description}
\begin{figure}[!t]\centering
	\includegraphics[width=11cm]{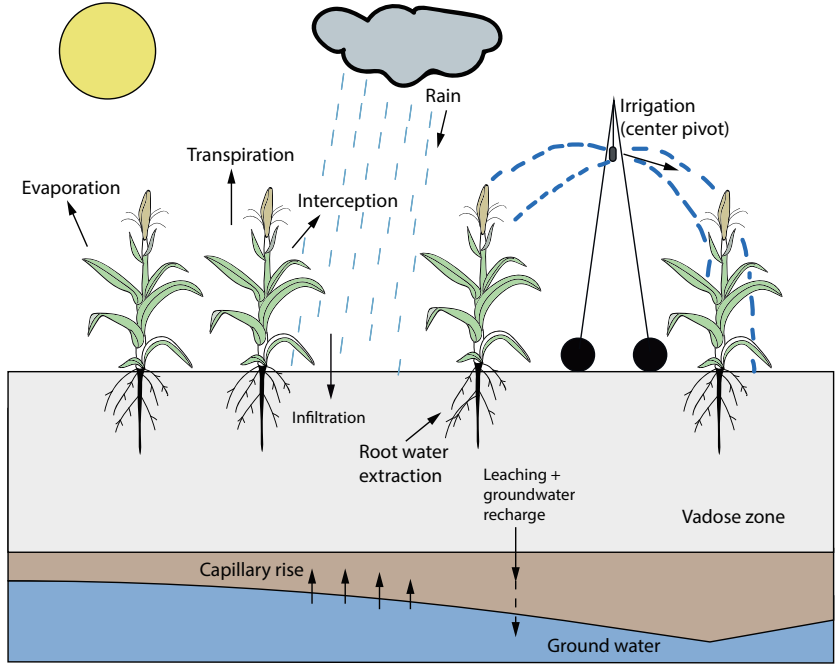}
	\caption{A schematic of the considered agro-hydrological system \cite{bo2020parameter}} \label{Agro}
\end{figure}
This study considers the hydrological cycle of an agro-hydrological system, which involves the interactions between the soil, plants, and atmosphere. The water flows that enter and leave the soil, such as rainfall, irrigation, root absorption, evaporation, drainage, and runoff, are all considered as inputs to this system. Figure \ref{Agro} provides a schematic of the system under consideration. This study only focuses on the vadose zone of the field. Additionally, two hypotheses are formed: (a) the horizontal homogeneity of the soil's features; and (b) the uniformity of the field's surface irrigation. To comprehensively analyze the hydrological processes in the soil, this study considers the vertical dynamics of the system. Specifically, the dynamics of the water flows are described by the Richards equation \cite{richards1931capillary}:
\begin{eqnarray}\label{Richards} 
	\frac{\partial \theta}{\partial t}=C(h) \frac{\partial h}{\partial t}=\frac{\partial}{\partial z}\left[K(h)\left(\frac{\partial h}{\partial z}+1\right)\right]-S\left(h,z\right)
\end{eqnarray}

\begin{table}
	\caption{Parameters of Richards equation\label{parameter}}
	\centering
	% \newcolumntype{C}{>{\centering\arraybackslash}X}
	\begin{tabular}{ccc}
	\hline
	\textbf{Parameter}	& \textbf{Explanation}	& \textbf{Unit}\\
	\hline
	$\theta$		&moisture content			& $m^3/m^3$\\
    $z$				&vertical position			& $m$\\
	$h$				&capillary pressure head	& $m$\\
	$K(h)$          &unsaturated hydraulic conductivity & $m/s$\\
	$C(h)$			&capillary capacity			& $m^{-1}$\\
	$S(h,z)$        &root water extraction rate \cite{feddes1978simulation} & $m^3m^{-3}s^{-1}$\\
	$K_{s}(h)$      &saturated hydraulic conductivity &$m/s$ \\
	$\theta_{s}$    &saturated moisture content  & $m^3/m^3$\\
	$\theta_{r}$    &residual moisture content   & $m^3/m^3$\\
	\hline
	\end{tabular}
\label{table}
\end{table}

The soil hydraulic properties, including the hydraulic conductivity $K(h)$ and the capillary capacity $C(h)$, can be modeled using the van Genuchten-Mualem approach, which relates these properties to the soil water pressure head $h$. The expressions for $K(h)$ and $C(h)$ are given as follows:
\begin{eqnarray}\label{Conductivity} 
	K(h)=K_{s}\left[\left(1+(-\alpha h)^{n}\right)^{-\left(1-\frac{1}{n}\right)}\right]^{\frac{1}{2}}\left[1-\left[1-\left[\left(1+(-\alpha h)^{n}\right)^{-\left(1-\frac{1}{n}\right)}\right]^{\frac{n}{n-1}}\right]^{1-\frac{1}{n}}\right]^{2}, 
\end{eqnarray}
\begin{eqnarray}\label{capacity} 
	c(h)=\left(\theta_{s}-\theta_{r}\right) \alpha n\left(1-\frac{1}{n}\right)(-\alpha h)^{n-1}\left[1+(-\alpha h)^{n}\right]^{-\left(2-\frac{1}{n}\right)},
\end{eqnarray}

The explanations of the variables in equations (\ref{Richards}) to (\ref{capacity}) are listed in Table \ref{table}. The $\alpha (1/m)$ and $n$ are curve-fitting soil hydraulic properties. The Mualem-van Genuchten model \cite{van1980closed} represents the relationship between the pressure head $h$ and soil moisture $\theta$ as follows:
\begin{eqnarray}\label{Content} 
	\theta(h)=\left(\theta_{s}-\theta_{r}\right)\left[\frac{1}{1+(-\alpha h)^{n}}\right]^{1-\frac{1}{n}}+\theta_{r},
\end{eqnarray} 

Since the Richards equation is a partial differential equation, it does not have a closed-form solution. Therefore, in this study, the finite difference method is used to approximate the equation. The derivatives of equation (\ref{Richards}) with respect to the vertical property z are discretized using the finite difference method \cite{agyeman2021soil}. This approximation transforms the partial differential equation into a system of ordinary differential equations that depends solely on the temporal variable. Subsequently, the Euler method is used to approximate the temporal differential.

The boundary condition to solve the equation (\ref{Richards}) are illustrated below:
\begin{eqnarray}\label{Boun1} 
	\left.\frac{\partial h}{\partial z}\right|_T=-1-\frac{q_T}{K(h)},
\end{eqnarray}
\begin{eqnarray}\label{Boun2} 
	\left.\frac{\partial(h+z)}{\partial z}\right|_B=1.
\end{eqnarray}
The equations denote boundary conditions at the top and bottom layers, respectively. The $q_T (m/s)$ is the irrigation rate which is considered as the manipulated input of the system and free drainage boundary condition is imposed at the bottom.

The nonlinear state-space model is represented as follows:
\begin{eqnarray}\label{State1} 
	x_{k+1}=f(x_{k}, u_{k})+\omega_{k}
\end{eqnarray}
where $x_k$ denotes the state vector with dimension of $n$, at the defined time instant $k$. $u_k$ and $\omega_k$ denote the input and the process noise with zero mean and covariance of $Q$, respectively.

\subsection{Problem formulation}

In this work, we consider the presence of model mismatch, which can be described as unknown inputs added to the state equation. Therefore, the actual system can be modeled as follows:
\begin{eqnarray}\label{State2} 
	x_{k+1}=f(x_{k}, u_{k})+Ma_{k}+\omega_{k}
\end{eqnarray}
where $a_{k}$ is the vector of unknown inputs which represents the model mismatch, $M$ is an identity matrix, which means the unknown inputs will affect all the states.
The general output function that considers measurement noise can be expressed as follows:
\begin{eqnarray}\label{Meas1} 
	y_{k+1}=Ch(x_{k+1})+v_{k+1}
\end{eqnarray}
where $y_{k+1}$ and $v_{k+1}$ denote the $m$ dimensional measurement vector and measurement noise with zero mean and covariance $R$. The $C$ matrix represents the location of the sensor. In this work, the soil moisture sensor is used to measure the volumetric soil moisture $\theta$, the measurement equation (\ref{Meas1}) is converted from the equation (\ref{Content}). The main aim of this work is to develop a computationally efficient approach to estimate the states when there is model mismatch in the form of additive unknown inputs.

\section{Extended Kalman Filter Based Recursive EM Algorithm}
\label{EKF-REM}
Given the model formulated in the previous section, the objective is now to simultaneously estimate the states and unknown inputs to address the state estimation problem in agro-hydrological systems with model mismatch. To achieve this goal, the EM algorithm is commonly utilized to estimate the expectation of latent variables and unknown parameters. To overcome the complexities inherent in high-dimensional agro-hydrological systems, we modified the traditional EM algorithm to a recursive version. The following section outlines the detailed derivation of this approach and the following notation will be used throughout this section: $\{x_0, \cdots, x_N\}$ refers to the set of states, $\{y_1, \cdots, y_N\}$ represents the set of measurements, $a_{0:N}$ is the set of unknown inputs. $\mathbb{E}$ denotes the mathematical expectation, $\operatorname{Tr}$ is the trace of the matrix. The notation $N(\mu, P)$ represents a Gaussian probability distribution with mean $\mu$ and covariance $P$. The superscripts '$\wedge$' and '$\vee$' denote the estimation and prediction, respectively.

Firstly, we will discuss the traditional EM algorithm, also known as the batch EM (BEM) algorithm. The BEM algorithm aims to maximize the expected likelihood of complete data with respect to latent variables. In each iteration, the algorithm updates the parameters by optimizing the likelihood. In the E-step, the Q-function is typically computed, which represents the expected log-likelihood of complete data. The Q-function is expressed as follows:
\begin{eqnarray}\label{Qfun}
	Q\left(a, a^{\prime}\right)=\mathrm{E}_{z_{\text {mis }} \mid \left(z_{o b s}, a^{\prime}\right)}\left\{\log p\left(z_{\text {mis }}, z_{\text {obs }} \mid a\right)\right\}
\end{eqnarray}
where the $z_{mis}$ are the latent variables which means the states in this paper, $z_{obs}$ are the measurements. The log-likelihood can be formulated in terms of the state-space model's Markov-chain structure. Specifically, the log-likelihood function is expressed as follows:
\begin{eqnarray}\label{log-likelihood} 
	L_{0:N}&=& \log p\left(x_{0}, \ldots, x_{N}, y_{1}, \ldots, y_{N} \mid a_{0:N-1}\right) \nonumber \\&=& \log p\left(x_{0} \mid a_{0}\right)+\sum_{k=1}^{N} \log p\left(x_{k} \mid x_{k-1}, a_{k-1}\right)+\sum_{k=1}^{N} \log p\left(y_{k} \mid x_{k}\right)
\end{eqnarray}

During the M-step, the goal is to find the optimal solution for the unknown inputs by maximizing the Q-function, this yields the following expression:
\begin{eqnarray}\label{Max}
	a=\arg \max _{a} Q\left(a, a^{\prime}\right)
\end{eqnarray}

By iteratively performing the preceding two procedures, the Q-function can be locally maximized.

\subsection{Recursive Q-function in E-step}

The BEM algorithm is an optimization algorithm that requires all historical data to be considered during each iteration. However, this approach can be computationally complex and is not well-suited for real-time applications. An alternative to the BEM algorithm is the EM algorithm, which utilizes a recursive computation of the Q-function. By using this approach, the EKF-based REM strategy can be implemented for online state and unknown inputs estimation. The following derivation illustrates this concept in more detail.

To rewrite equation (\ref{log-likelihood}), we can exclude the cumulative items from the last time instance:
\begin{eqnarray}\label{Q-func2} 
	Q\left({a}, {a}^{\prime}\right)&=& \mathbb{E}\left\{\log p\left({x}_{0} \mid {a}_{0}\right)+\sum_{k=1}^{N-1} \log p\left({x}_{k} \mid {x}_{k-1}, {a}_{k-1}\right)\right. \nonumber \\ && \left.+\sum_{k=1}^{N-1} \log p\left({y}_{k} \mid {x}_{k}\right)+\log p\left({x}_{N} \mid {x}_{N-1}, {a}_{N-1}\right)+\log p\left({y}_{N} \mid {x}_{N}\right)\right\}
\end{eqnarray}

The REM algorithm updates equation (\ref{Q-func2}) as new data is received in a time sequence. Unlike the BEM approach, which involves all historical data, the REM algorithm performs a single E-step using only the latest data and parameters identified in the previous time index. This allows for a quasi-recursive formulation of equation (\ref{Q-func2}) that can be expressed as follows:
\begin{eqnarray}\label{quasi-Q-func1} 
	Q_{N}\left({a},{a}_{N}^{\text {old}}\right)&=&\mathbb{E}\left\{\log p\left({x}_{0} \mid {a}_{0}\right)+\sum_{k=1}^{N-1} \log p\left({x}_{k} \mid {x}_{k-1}, {a}_{k-1}\right)\right. \nonumber \\ && \left. +\sum_{k=1}^{N-1} \log p\left({y}_{k} \mid {x}_{k},\right)+\log p\left({x}_{N} \mid {x}_{N-1}, {a}_{N-1}\right)+\log p\left({y}_{N} \mid {x}_{N}\right)\right\}
\end{eqnarray}
the subscript $N$ represents the current time instant, and $a$ denotes the set of unknown inputs to be estimated in the current moment. The variable $a_N^{old}$ refers to the solution obtained at time index $N-1$, which is used to compute the posterior expectation at the $N$ time instant. The superscript 'old' indicates that the unknown inputs were estimated at the preceding time step $N$. The estimation of hidden states is achieved by utilizing the parameters that are recursively obtained in the quasi-recursive Q-function (\ref{quasi-Q-func1}). Consequently, the quasi-recursive Q-function can be mathematically expressed as follows:
\begin{eqnarray}\label{quasi-Q-func2} 
	Q_{N}\left({a}, a_{N}^{\text {old}}\right)=Q_{N-1}\left({a}, {a}_{N-1}^{\text {old}}\right)+\mathbb{E}\left\{\log p\left({x}_{N} \mid {x}_{N-1}, {a}_{N-1}\right)+\log p\left({y}_{N} \mid {x}_{N}\right)\right\}
\end{eqnarray}

Equation (\ref{quasi-Q-func2}) can be categorized as a quasi-recursion because the E-step is independent of the time index. A true recursive Q-function can be introduced by the definition of $\tilde{Q}_{N}$, which is the average of the ${Q}_{N}$ at time index $N$:
\begin{eqnarray}\label{quasi-Q-func3} 
	\tilde{Q}_{N}\left({a}, {a}_{N}^{\text {old }}\right) = \frac{1}{N} Q_{N}\left({a}, {a}_{N}^{\text {old }}\right) 
\end{eqnarray}

After substituting equation (\ref{quasi-Q-func3}) into the above equation (\ref{quasi-Q-func2}), we can obtain the recursive form for the likelihood of complete data, as shown below:
\begin{eqnarray}\label{quasi-Q-func4} 
	\tilde{Q}_{N}\left({a}, {a}_{N}^{\text {old }}\right) &=& \frac{1}{N} Q_{N-1}\left({a}, {a}_{N-1}^{\text {old }}\right) +\frac{1}{N} \mathbb{E}\left\{\log p\left({x}_{N} \mid {x}_{N-1}, {a}_{N-1}\right)\nonumber+\log p\left({y}_{N} \mid {x}_{N}\right)\right\}\nonumber \\ &=& (1-\frac{1}{N}) \tilde{Q}_{N-1}\left({a}, {a}_{N-1}^{\text {old }}\right)+\frac{1}{N} \mathbb{E}\left\{\log p\left({x}_{N} \mid {x}_{N-1}, {a}_{N-1}\right)\right.\nonumber \\ && \left.+\log p\left({y}_{N} \mid {x}_{N}\right)\right\}
\end{eqnarray}

The natural step-size is decreasing with the time index, it may not always be the most effective approach. To handle this problem, we introduce a fixed step-size or the learning rate as a hyper-parameter which can be tuned to replace $1/N$. From the theory of stochastic approximation \cite{chen2020online}, we employ an artificial step-size denoted by $\gamma_{N}$ to replace the step-size $1/N$ in (\ref{quasi-Q-func4}). Consequently, the expectation can be recursively calculated as:
\begin{eqnarray}\label{rec-Q-func} 
	\tilde{Q}_{N}\left({a}, {a}_{N}^{\text {old }}\right)=(1-\gamma_{N}) \tilde{Q}_{N-1}\left({a}, {a}_{N-1}^{\text {old }}\right)+\gamma_{N} \mathbb{E}\left\{\log p\left({x}_{N} \mid {x}_{N-1}, {a}_{N-1}\right)+\log p\left({y}_{N} \mid {x}_{N}\right)\right\}
\end{eqnarray}

By computing the recursion of $\tilde{Q}_{N}\left({a}, {a}_{N}^{\text {old }}\right)$ from the initial time instance, the recursive Q-function (\ref{rec-Q-func}) is further elaborated as follows:
\begin{eqnarray}\label{rec-Q-func2} 
	\tilde{Q}_{N}\left(a, a_{N}^{\text {old }}\right)= \prod_{t=2}^{N}\left(1-\gamma_{t}\right)\mathbb{E}(G_{1})+
	\sum_{k=2}^{N-1}\left[\prod_{t=k+1}^{N}\left(1-\gamma_{t}\right)\right]\gamma_{k}\mathbb{E}(G_{k})+\gamma_{N}\mathbb{E}(G_{N})
\end{eqnarray}
where
\begin{eqnarray}\label{EG1}
	\mathbb{E}\left(G_{1}\right)= \mathbb{E}\left\{\log {p}\left({x}_{0} \mid {a}_{0}\right)+\log {p}\left({x}_{1} \mid {x}_{0}, {a}_{0}\right)+\log {p}\left({y}_{1} \mid {x}_{1} \right) \right\}
\end{eqnarray}
\begin{eqnarray}\label{EGk} 
	\mathbb{E}\left(G_{k}\right) = \mathbb{E}\left\{\log {p}\left({x}_{k} \mid {x}_{k-1}, {a}_{k-1}\right) +\log {p}\left({y}_{k} \mid {x}_{k}\right) \right\}
\end{eqnarray}
\begin{eqnarray}\label{EGN}
	\mathbb{E}\left(G_{N}\right)=\mathbb{E}\left\{\log {p}\left({x}_{N} \mid {x}_{N-1}, {a}_{N-1}\right)+\log {p}\left({y}_{N} \mid {x}_{N}\right) \right\}
\end{eqnarray}

To calculate equations (\ref{EG1})-(\ref{EGN}), the conditional probablity density functions of initial guess ${p}\left({x}_{0} \mid {a}_{0}\right)$, state update ${p}\left({x}_{k} \mid {x}_{k-1}, {a}_{k-1}\right)$, and measurements ${p}\left({y}_{k} \mid {x}_{k}\right)$ are required. Under the assumption that the system noise and measurement noise are both Gaussian, the abovementioned equations are represented as below
\begin{eqnarray}\label{gaussian1}
	{p}\left({x}_{0} \mid {a}_{0}\right)={N}\left(\hat{{x}}_{0}, {P}_{0}\right) 
\end{eqnarray}
\begin{eqnarray}\label{gaussian2}
	\begin{aligned}
		{p}\left({x}_{k} \mid {x}_{k-1}, {a}_{k-1}\right)={N}\left({f} ({x}_{k-1}, {u}_{k-1})+{M} {a}_{k-1}, {Q}_{k}\right)
	\end{aligned}
\end{eqnarray}
\begin{eqnarray}\label{gaussian3}
	{p}\left({y}_{k} \mid {x}_{k}\right)={N}\left({C}{h}({x}_{k}), {R}_{k}\right) 
\end{eqnarray}

Next, we can derive the log-likelihood of the aforementioned equations (\ref{gaussian1})-(\ref{gaussian3}) as follows:
\begin{eqnarray}\label{log1}
	\log {p}\left({x}_{0} \mid {a}_{0}\right)=-\frac{n}{2} \log (2 \pi)-\frac{1}{2} \log \left|{P}_{0}\right|-\frac{1}{2}\mathscr{D}\left( {x}_{0}-\hat{{x}}_{0},{P}_{0}\right)
\end{eqnarray}
\begin{eqnarray}\label{log2}
	\log {p}\left({x}_{k} \mid {x}_{k-1}, {a}_{k-1}\right)=-\frac{n}{2} \log (2 \pi)-\frac{1}{2} \log \left|{Q}_{k}\right|\frac{1}{2}\mathscr{D}\left( {x}_{k}-{f}\left({x}_{k-1},{u}_{k-1}\right)-{M}{a}_{k-1},{Q}_{k}\right)
\end{eqnarray}
\begin{eqnarray}\label{log3}
	\log {p}\left({y}_{k} \mid {x}_{k}\right)=-\frac{m}{2} \log (2 \pi)-\frac{1}{2} \log \left|{R}_{k}\right|-\frac{1}{2}\mathscr{D}\left( {y}_{k}-{C}{h}({x}_{k}),{R}_{k}\right) 
\end{eqnarray}
where the symbol $\mathscr{D}$ denotes for the abbreviation of matrix operation $\mathscr{D}(x,P)=x^{\operatorname{T}}P^{-1}x$.

Then we can obtain the expectation of the likelihood functions (\ref{log1})-(\ref{log3}):
\begin{eqnarray}\label{E1}
	\mathbb{E}\left( \log {p}\left({x}_{0} \mid {a}_{0}\right)\right)=-\frac{n}{2} \log (2 \pi)-\frac{1}{2} \log \left|{P}_{0}\right|-\frac{n}{2}
\end{eqnarray}
\begin{eqnarray}\label{E2}
	\begin{aligned}
		\mathbb{E}\left( \log {p}\left({x}_{k} \mid {x}_{k-1}, {a}_{k-1}\right)\right) =&-\frac{n}{2} \log (2 \pi)-\frac{1}{2} \log \left|{Q}_{k}\right|-\frac{1}{2}\operatorname{Tr}\Big\{ {Q}_{k}^{-1}\Big\{\mathscr{C}\left(\hat{{x}}_{k}\right.  \\ &\left.-{f} \left(\hat{{x}}_{k-1}, {u}_{k-1}\right) -{M}{a}_{k-1}\right)+\left( \hat{{P}}_{k}-{F_k}\check{{P}}_{k} \right) \\ &-\left(\check{{P}}_{k}{F_k}^{\mathrm{T}}-{F_{k-1}}\hat{{P}}_{k-1}{F_{k-1}}^{\mathrm{T}} \right) \Big\}\Big\}
	\end{aligned}
\end{eqnarray}
\begin{eqnarray}\label{E3}
	\begin{aligned}
		\mathbb{E}\left( \log {p}\left({y}_{k} \mid {x}_{k}\right)\right) =&-\frac{m}{2} \log (2 \pi)-\frac{1}{2} \log \left|{R}_{k}\right|\\ &-\frac{1}{2} \operatorname{Tr}\Big\{{R}_{k}^{-1}\Big[\mathscr{C}\left({y}_{k}-C{H_k}\hat{{x}}_{k}\right)+\left(C{H_k}\hat{{P}}_{k}{H_k}^{\mathrm{T}}{C}^{\mathrm{T}}\right)\Big]\Big\}
	\end{aligned}
\end{eqnarray}
where $F_k$ and $H_k$ are the Jacobian matrices of $f(\cdot)$ and $h(\cdot)$ at time step $k$, respectively. The symbol $\mathscr{C}$ denotes for the abbreviation of matrix operation $\mathscr{C}(x)=xx^{\operatorname{T}}$. Then, we can 
\begin{eqnarray}\label{E(G1)} 
	\mathbb{E}\left(G_{1}\right) &=&-\frac{n+(m+n)}{2} \log (2 \pi)-\frac{1}{2}(\log \left|{Q}_{0}\right|+\log\left|{R}_{0}\right|)-\frac{1}{2} \log \left|\hat{{P}}_{0}\right|-\frac{n}{2}\nonumber \\ &&-\frac{1}{2} \operatorname{Tr}\Big\{{Q}_{1}^{-1}\Big\{\mathscr{C}\left(\hat{{x}}_{1}-{f}({\hat{x}_0},{u_0})-{M} {a}_{0}\right)+\left(\hat{{P}}_{1}-{F}_{1} \check{{P}}_{1}\right)\nonumber \\ &&-\left(\check{{P}}_{1} {F}_{1}^{\mathrm{T}}-{F}_{0} \hat{{P}}_{0} {F}_{0}^{\mathrm{T}}\right)\Big\}\Big\} -\frac{1}{2} \operatorname{Tr}\left\{{R}_{1}^{-1} \left\{\left(C{H}_{1} \hat{{P}}_{1} {H}_{1}^{\mathrm{T}}C^{\mathrm{T}}\right) +\mathscr{C}\left({y}_{1}-C{h} (\hat{{x}}_{1})\right)\right\}\right\}
\end{eqnarray}
\begin{eqnarray}\label{E(Gk)} 
	\mathbb{E}\left(G_{k}\right)&=&-\frac{m+n}{2} \log (2 \pi)-\frac{1}{2}(\log \left|{Q}_{k}\right|+\log\left|{R}_{k}\right|)-\frac{1}{2} \log \left|\hat{{P}}_{k}\right|-\frac{1}{2} \operatorname{Tr}\Big\{{Q}_{k}^{-1} \nonumber \\ && \Big\{\mathscr{C}\big(\hat{{x}}_{k}-{f} (\hat{{x}}_{k-1}, {u}_{k-1})-{M} {a}_{k-1}\big)+\Big(\hat{{P}}_{k}-{F}_{k} \check{{P}}_{k}\Big) -\Big(\check{{P}}_{k} {F}_{k}^{\mathrm{T}}-{F}_{k-1} \hat{{P}}_{k-1} {F}_{k-1}^{\mathrm{T}}\Big)\Big\}\Big\}\nonumber \\ &&-\frac{1}{2} \operatorname{Tr}\Big\{{R}_{k}^{-1} \Big\{\Big({H}_{k} \hat{{P}}_{k} {H}_{k}^{\mathrm{T}}\Big)+\mathscr{C}\big({y}_{k}-C{h} (\hat{{x}}_{k})\big)\Big\}\Big\}
\end{eqnarray}
\begin{eqnarray}\label{E(GN)} 
	\mathbb{E}\left(G_{N}\right)&=&-\frac{m+n}{2} \log (2 \pi)-\frac{1}{2}(\log \left|{Q}_{N}\right|+\log\left|{R}_{N}\right|)-\frac{1}{2} \log \left|\hat{{P}}_{N}\right|-\frac{1}{2} \operatorname{Tr}\Big\{\nonumber \\ &&{Q}_{N}^{-1}\Big\{\mathscr{C}\big(\hat{{x}}_{N}-{f}(\hat{{x}}_{N-1},{u}_{N-1})-{M} {a}_{N-1}\big)+\left(\hat{{P}}_{N}-F_{N-1} \check{{P}}_{N}\right)\nonumber \\ &&-\left(\hat{{P}}_{N} {F}_{N}^{\mathrm{T}}-{F}_{N-1}\hat{{P}}_{N-1} {F}_{N-1}^{\mathrm{T}}\right)\Big\}\Big\}-\frac{1}{2} \operatorname{Tr}\Big\{{R}_{N}^{-1}\Big\{\left(C{H}_{N} \hat{{P}}_{N} {H}_{N}^{\mathrm{T}}C^{\mathrm{T}}\right)\nonumber \\ &&+\mathscr{C}\left({y}_{N}-C{h}(\hat{{x}}_{N})\right)\Big\}\Big\}
\end{eqnarray}
where $\prod_{t=k+1}^{N}\left(1-\gamma_{t}\right)$ is the product of the step-sizes from time $k+1$ to $N$, $t$ is the corresponding time index. The derivation of the recursive Q-function is completed.

\begin{algorithm}
	\caption{EKF based Recursive Expectation-Maxization Algorithm}
	\label{alg1}
	\begin{algorithmic}[1]
		\renewcommand{\algorithmicrequire}{\textbf{Data:}}
		\renewcommand{\algorithmicensure}{\textbf{Result:}}
		\REQUIRE ${{y}}_N$
		\ENSURE  ${\hat {{x}}_{N}}$, ${{a}}_{N}$
		\\
		\STATE Initialize sufficient statistics and unknown inputs: $a_{1}$
		\FOR {$N = 2:k$}
		\STATE \textbf{R E-step} Update sufficient statistics:
		\\
		$\tilde{Q}_{N}\left({a}, {a}_{N}^{\text {old }}\right)=(1-\gamma_{N}) \tilde{Q}_{N-1}\left({a}, {a}_{N-1}^{\text {old }}\right)+\gamma_{N} \mathbb{E}\left\{\log p\left({x}_{N} \mid {x}_{N-1}, {a}_{N-1}\right)\nonumber +\log p\left({y}_{N} \mid {x}_{N}\right)\right\}$
		\\
		\STATE \textbf{Prediction}
		\\
		$\check{{x}}_{N} ={f}\left(\hat{{x}}_{N-1},{u}_{N-1}\right)+{M}{a}_{N-1}$
		\\
		$\check{{P}}_{N} ={F}_{N-1}\hat{{P}}_{N-1}{F}_{N-1}^{\mathrm{T}}+{Q}_{N-1}$
		\\
		\STATE \textbf{Update}
		\\
		${K}_{N} =\check{{P}}_{N}{H}_{N}^{\mathrm{T}}C^{\mathrm{T}}\left(C{H}_{N}\check{{P}}_{N} {H}_{N}^{\mathrm{T}}C^{\mathrm{T}}+{R}_{N}\right) ^{-1}$
		\\
		$\hat{{x}}_{N}=\check{{x}}_{N}+{K}_{N}\left( {y}_{N}-C{h}\left(\check{{x}}_{N}\right)\right)$
		\\
		$\hat{{P}}_{N}=\left({I}-{K}_{N} {H}_{N}\right) \check{{P}}_{N} $
		\\
		\STATE \textbf{M-step} Maximize the expectaion: $\frac{\partial \tilde{Q}_{N}\left({a},{a}_{N}^{\text {old }}\right)}{\partial {a}_{N}}=0$
		\\
		${a}_{N}=\left(1-\gamma_{N}\right){a}_{N-1}+\gamma_{N}\left( {M}^{-1}\left(\hat{{x}}_{N}-{f}\left(\hat{{x}}_{N-1}, {u}_{N-1}\right) \right) \right)$	
		\ENDFOR
		\\
	\end{algorithmic}
\end{algorithm}

\subsection{Predict and Update of States and Covariance}

To compute the conditional expectation of the log-likelihood function (\ref{rec-Q-func2}), it is necessary to evaluate the posterior probability and prediction of ${x}_{N}$ and corresponding covariance ${P}_{N}$. Previous studies have employed a fixed-interval smoother to calculate the posterior of the hidden states in the Q-function \cite{lan2013joint,khan2019simultaneous}. While smoothers are effective batch processing methods for state estimation, they are not suitable for real-time computation. Therefore, in this study, we utilize the extended Kalman filter (EKF) for state estimation, which can be computed recursively. The EKF is capable of providing sub-optimal state estimation for nonlinear stochastic systems. Its recursive calculation capability makes it compatible with integration into the REM algorithm. The filter can be represented as follows:

Predicted state estimation:
\begin{eqnarray}\label{27}
	\check{{x}}_{N} ={f}\left(\hat{{x}}_{N-1},{u}_{N-1}\right)+{M}{a}_{N-1}
\end{eqnarray}

Predicted covariance estimation:
\begin{eqnarray}\label{28} 
	\check{{P}}_{N} ={F}_{N-1}\hat{{P}}_{N-1}{F}_{N-1}^{\mathrm{T}}+{Q}_{N-1}
\end{eqnarray}

Near-optimal Kalman gain:
\begin{eqnarray}\label{29} 
	{K}_{N} =\check{{P}}_{N}{H}_{N}^{\mathrm{T}}C^{\mathrm{T}}\left(C{H}_{N}\check{{P}}_{N} {H}_{N}^{\mathrm{T}}C^{\mathrm{T}}+{R}_{N}\right) ^{-1}
\end{eqnarray}

Updated state estimation:
\begin{eqnarray}\label{30} 
	\hat{{x}}_{N}=\check{{x}}_{N}+{K}_{N}\left({y}_{N}-C{h}\left(\check{{x}}_{N}\right)\right) 
\end{eqnarray}

Updated covariance estimation:
\begin{eqnarray}\label{31} 
	\hat{{P}}_{N}=\left({I}-{K}_{N} {H}_{N}\right) \check{{P}}_{N}
\end{eqnarray}
where $F_N$ and $H_N$ are the Jacobian matrices of $f(\cdot)$ and $h(\cdot)$ at time step $N$, respectively.
\subsection{M-step}

To achieve the maximum solution of the recursive Q-function in the EKF-based REM algorithm, we perform partial differentiation with respect to the unknown inputs. By setting the partial derivative to zero, we can calculate the local maximum of the model's unknown input vector $a_{k}$.
\begin{eqnarray}\label{32} 
	\frac{\partial \tilde{Q}_{N}\left({a},{a}_{N}^{\text {old }}\right)}{\partial {a}_{N}}=0
\end{eqnarray}

Then, by substituting equation (\ref{rec-Q-func2}) to equation (\ref{32}), we can obtain
\begin{eqnarray}\label{34} 
	\frac{\partial \tilde{Q}_{N}\left({a}, {a}_{N}^{\text {old }}\right)}{\partial {a}_{N}}&=&\frac{\partial}{\partial {a}_{N}}\Bigg\{ \prod_{t=2}^{N}\left(1-\gamma_{t}\right)\mathbb{E}(G_{1})+\sum_{k=2}^{N-1}\left[\prod_{t=k+1}^{N}\left(1-\gamma_{t}\right)\right]\gamma_{k}\mathbb{E}(G_{k})\nonumber \\ &&+\gamma_{N}\mathbb{E}(G_{N})\Bigg\}=0
\end{eqnarray}

Finally, we can get
\begin{eqnarray}\label{M-step} 
	{a}_{N}&=&\left(1-\gamma_{N}\right){a}_{N-1}+\gamma_{N}\left( {M}^{-1}\left(\hat{{x}}_{N}-{f}\left(\hat{{x}}_{N-1}, {u}_{N-1}\right) \right) \right) 
\end{eqnarray}

Therefore, the derivation of the M-step of EKF based REM algorithm is completed. The pseudocode of the proposed EKF based REM approach is illustrated in Algorithm \ref{alg1}.

\section{Sensor placement}
\label{Sensor}
To reduce the cost and maintenance associated with estimating all of the system's states using sensors in the 1D agro-hydrological system, it is possible to choose a minimum number of sensors that can still guarantee the observability of the system. This can be achieved by constructing a local sensitivity matrix that measures the sensitivity of the system's outputs to its initial conditions, it can help determine the extent to which a system can be observable locally. In \cite{liu2021simultaneous}, a criterion based on the local sensitivity matrix is proposed to assess the degree of observability for different sensor combinations, using the successive orthogonalization project. To select an optimal subset of sensors that can provide reliable estimates for the states of the agro-hydrological system, a ranking of sensor relevance is established based on the local sensitivity matrix. After constructing the matrix, the optimal subset of sensors can be obtained by selecting the state nodes that are least relevant while ensuring that the matrix maintains full column rank. This guarantees observability of the system. By using this method, it is possible to estimate the system states in a cost-effective way while still maintaining observability.

\subsection{Sensitivity Analysis}

For the agro-hydrological dynamic system, it is necessary to estimate all states and unknown inputs simultaneously. To achieve this, the sensitivity matrix must include both the states and unknown inputs. One commom approach to check the observability of the states and unknown inputs is augmenting the unknown inputs as states. For clarity and simplicity, in this section, we will use the subscript $a$ to represent the augmented system and $k$ to denote the time step. Therefore, the following augmented system can be obtained:
\begin{eqnarray}\label{Augmented} 
	{x_{\rm{a}}}(k + 1) &=& \left[ {\begin{array}{*{20}{l}}
	{f(x(k),u(k))}\\
	{a(k)}
	\end{array}} \right]: = {f_{\rm{a}}}\left( {{x_{\rm{a}}}(k),u(k)} \right)\\
	\nonumber y(k) &=& Ch(x(k)): = {C_a}{h_{\rm{a}}}\left( {{x_{\rm{a}}}(k)} \right)
\end{eqnarray}
where $x_{\mathrm{a}}(k)=\left[x^{\mathrm{T}}(k)\ a^{\mathrm{T}}(k)\right]^{\mathrm{T}} \in R^{n+p}$ denotes the augmented state vector. The rank of the successive sensitivity matrix along a typical trajectory within a data window of the augmented system is evaluated to assess the observability of the augmented system.
\begin{eqnarray}\label{Sensi} 
{S_{\rm{a}}}(k) = \left[ {\begin{array}{*{20}{l}}
	{{S_{y,{x_{\rm{a}}}(k - N)}}(k - N)}\\
	{{S_{y,{x_{\rm{a}}}(k - N)}}(k - N + 1)}\\
	 \vdots \\
	{{S_{y,{x_{\rm{a}}}(k - N)}}(k)}
	\end{array}} \right]
\end{eqnarray}
where $N$ is the size of data window and either larger than or equal to $n+p$.

Given state and measurement equations of a process, we can derive the following formula to determine sensitivity of the measurements $y(i),i=k-N,\cdots,k$ to the states $x_a(k-N)$ as
\begin{eqnarray}\label{Sensi2} 
	S_{y, x_{\mathrm{a}}(k-N)}(i)= C_{a}\frac{\partial H_{\mathrm{a}}}{\partial x_{\mathrm{a}}}(i) \frac{\partial F_{\mathrm{a}}}{\partial x_{\mathrm{a}}}(i-1) \frac{\partial F_{\mathrm{a}}}{\partial x}(i-2)  \cdots \frac{\partial F_{\mathrm{a}}}{\partial x_{\mathrm{a}}}(k-N)
\end{eqnarray}
where $\frac{\partial F_{\mathrm{a}}}{\partial x_{\mathrm{a}}(i)}=\left[\begin{array}{ll}
	\frac{\partial F}{\partial x}(i) & \frac{\partial F}{\partial a} \\
	0 & I_{p \times p}
	\end{array}\right], \quad \frac{\partial H_{\mathrm{a}}}{\partial x_{\mathrm{a}}}(i)=\left[\begin{array}{ll}
	\frac{\partial H}{\partial x}(i) & \frac{\partial H}{\partial a}
	\end{array}\right]$.
	
Examining the rank of the matrix $S_{a}(k)$ at each sampling period can help determine whether the complete augmented state vector $x_a$ can be estimated locally using input and measurement data. By evaluating the rank of the sensitivity matrix, we can determine whether the system is observable and if it is possible to estimate all variables accurately.

\subsection{Sensor Selection for Agro-hydrological system}

\begin{figure}[!t]\centering
	\includegraphics[width=12cm]{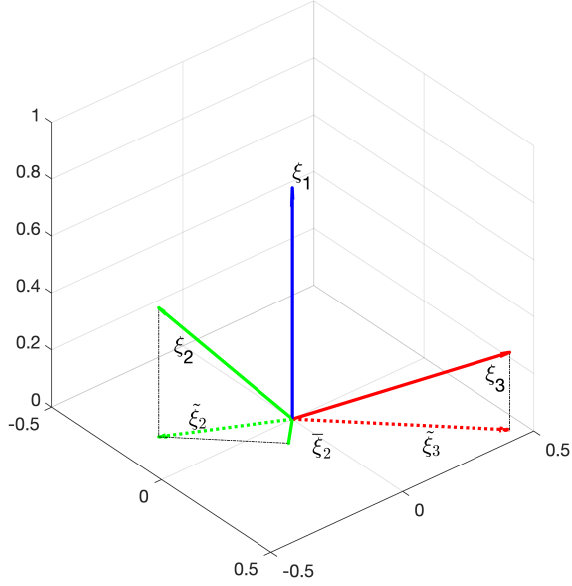}
	\caption{A schematic diagram of the  projection procedure}\label{proj}
\end{figure}

To determine the optimal sensor subset for agro-hydrological systems, the sensitivity matrix $S_{a}(k)$ needs to be analyzed for both the norm of columns and the degree of dependence between them. When the columns of a matrix are highly correlated, it suggests that changing the corresponding states would result in similar effects on the system's measurements. Successive orthogonalization is used to rank the columns of $S_{a}(k)$ \cite{liu2022sensor}. To perform the orthogonalization, the projection method is applied,  which is demonstrated in the following example using Algorithm \ref{Sensor1}. A schematic diagram of the orthographic projection described in the algorithm is shown in Figure \ref{proj}.

To reduce computational complexity, we adopt Algorithm \ref{Sensor1} for orthographic projection. The following is the procedure for sensor selection. Firstly, we select the column in the sensitivity matrix with the largest norm of sensitivity and place a sensor on the corresponding state. Then we calculate the rank of the sensitivity matrix. If it is full rank, it means that we only need one sensor to estimate all the states and unknown inputs. If it is not full rank, then we add a sensor to the state that has the largest component orthogonal to the selected one. We then recalculate the rank of the sensitivity matrix constructed by these two measurements. By repeating these two steps to ensure that the sensitivity matrix is full rank, which gives us the optimal and minimum set of sensors.

\begin{algorithm}
	\caption{Orthogonal Projection Algorithm}
	\label{Sensor1}
	\begin{algorithmic}[1]
		\STATE Construct the sensivity matrix ${S}_{a}(k):=[{\xi}_{1}, {\xi}_{2}, \cdots, {\xi}_{m}]$
		\STATE Calculate the norm of each column in matrix ${S}_{a}(k)$, if the $\xi_{1}$ has the largest norm $G_{\xi_{1}}$, then calculate the direction $p_{{\xi}_{1}}={\xi}_{1} / \Vert {\xi}_{1}\Vert$
		\STATE Calculate $\tilde{{\xi}}_{2}={\xi}_{2}-\left({p}_{{\xi}_{1}}^{\mathrm{T}} {\xi}_{2}\right) {p}_{{\xi}_{1}}$, $\tilde{{\xi}}_{3}={\xi}_{3}-\left({p}_{{\xi}_{1}}^{\mathrm{T}} {\xi}_{3}\right) {p}_{{\xi}_{1}}$ to $\tilde{{\xi}}_{m}={\xi}_{m}-\left({p}_{{\xi}_{1}}^{\mathrm{T}} {\xi}_{m}\right) {p}_{{\xi}_{1}}$, if $\tilde{{\xi}}_{3}$ has a larger norm than other columns, then calculate the direction ${p}_{\tilde{{\xi}}_{3}}=\tilde{{\xi}}_{3} / \Vert \tilde{{\xi}}_{3}\Vert$
		\STATE Repeat step 3 until end, based on the multiple orthogonalizations, the norm of each column which represents the irrelevance is arranged as $F_{{\xi}_{1}}>F_{{\xi}_{3}}>\cdots>F_{{\xi}_{m}}$

	\end{algorithmic}
\end{algorithm}

\section{Simulations}

\subsection{System description}

\begin{figure}[!t]\centering
	\includegraphics[width=6cm]{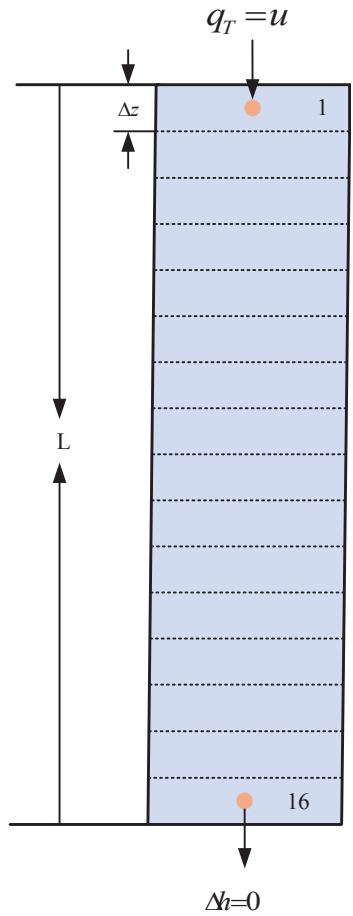}
	\caption{A diagram depicting the loam soil column under investigation and sensor locations}\label{Loam}
\end{figure}

In this case study, we investigate a soil column with a depth of $0.30 \ m$, as exhibited in Figure \ref{Loam}. The soil column is divided into $16$ compartments. By using the sensor placement method mentioned in Section \ref{Sensor}, we determine that only $2$ sensors are needed for this soil column, and the optimal locations for these sensors are indicated by the orange dots in Figure \ref{Loam}.

We assume that the soil column has the same capillary pressure head for all the $16$ states, and the soil parameters used in the model are provided in Table \ref{table1} \cite{carsel1988developing}.
\begin{table}[htbp]
	\centering
	\caption{The parameters of the investigated loam soil column.}
	\begin{tabular}{cccccc}
		\toprule & $K_s(\mathrm{m} / \mathrm{s})$ & $\theta_s\left(\mathrm{m}^3 / \mathrm{m}^3\right)$ & $\theta_r\left(\mathrm{m}^3 / \mathrm{m}^3\right)$ & $\alpha(\mathrm{1} / \mathrm{m})$ & $n$ \\
		\midrule Loam & $2.89 \times 10^{-6}$ & $0.430$ & $0.0780$ & $3.60$ & $1.56$ \\
		\bottomrule
	\end{tabular}
\label{table1}
\end{table}

In this section, we demonstrate the performance of the proposed EKF-based recursive EM algorithm with simulated measurements in three different scenarios:

(i) Scenario 1: The values of the unknown inputs are the same for each state, and the initial guesses of the unknown inputs are also the same.

(ii) Scenario 2: The values of the unknown inputs are different for each state, and the initial guesses of the unknown inputs are random.

(iii) Scenario 3: The unknown inputs are caused by errors in the crop efficiency $Kc$ and evaporation rate $Et$, which are involved in the sink term $S(h,z)$, and the two parameters are time-varying.

For each of these scenarios, we consider a sampling time of 2 min.

\subsection{Scenario 1}

In this scenario, all unknown inputs corresponding to each state have the same true value of $3 \times 10^{-5}$. The true value of the initial state is $x_{0}=-1$, and the initial guess of the state is $\hat{x}_{0}=1.1 \times x_{0}$. We consider process noise and measurement noise in this simulation, which have zero mean and variance of $4 \times 10^{-9}$ and $8 \times 10^{-7}$, respectively.

\begin{figure}[!t]\centering
	\includegraphics[width=11cm]{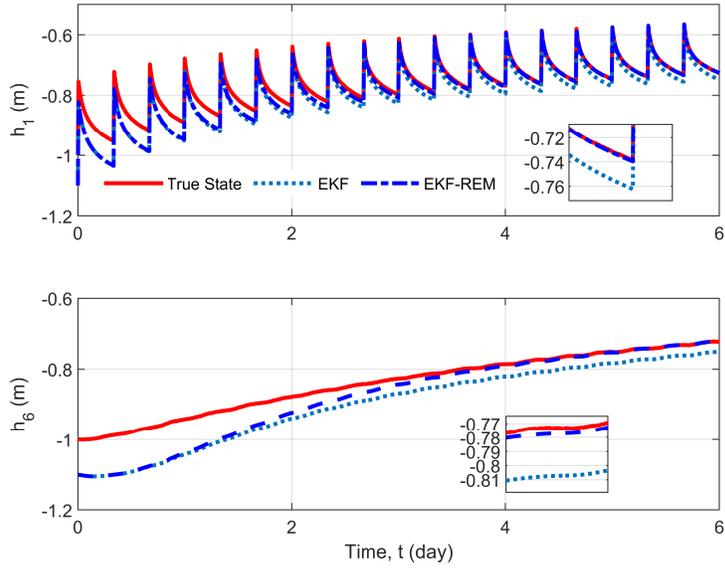}
	\caption{Trajectories of the actual state and the estimates by the standard EKF and the EKF-based REM methods at the surface $h_1$ and depth of $\boldsymbol{10}\ cm \ $ $h_6$.}\label{State_C1}
\end{figure}

\begin{figure}[!t]\centering
	\includegraphics[width=11cm]{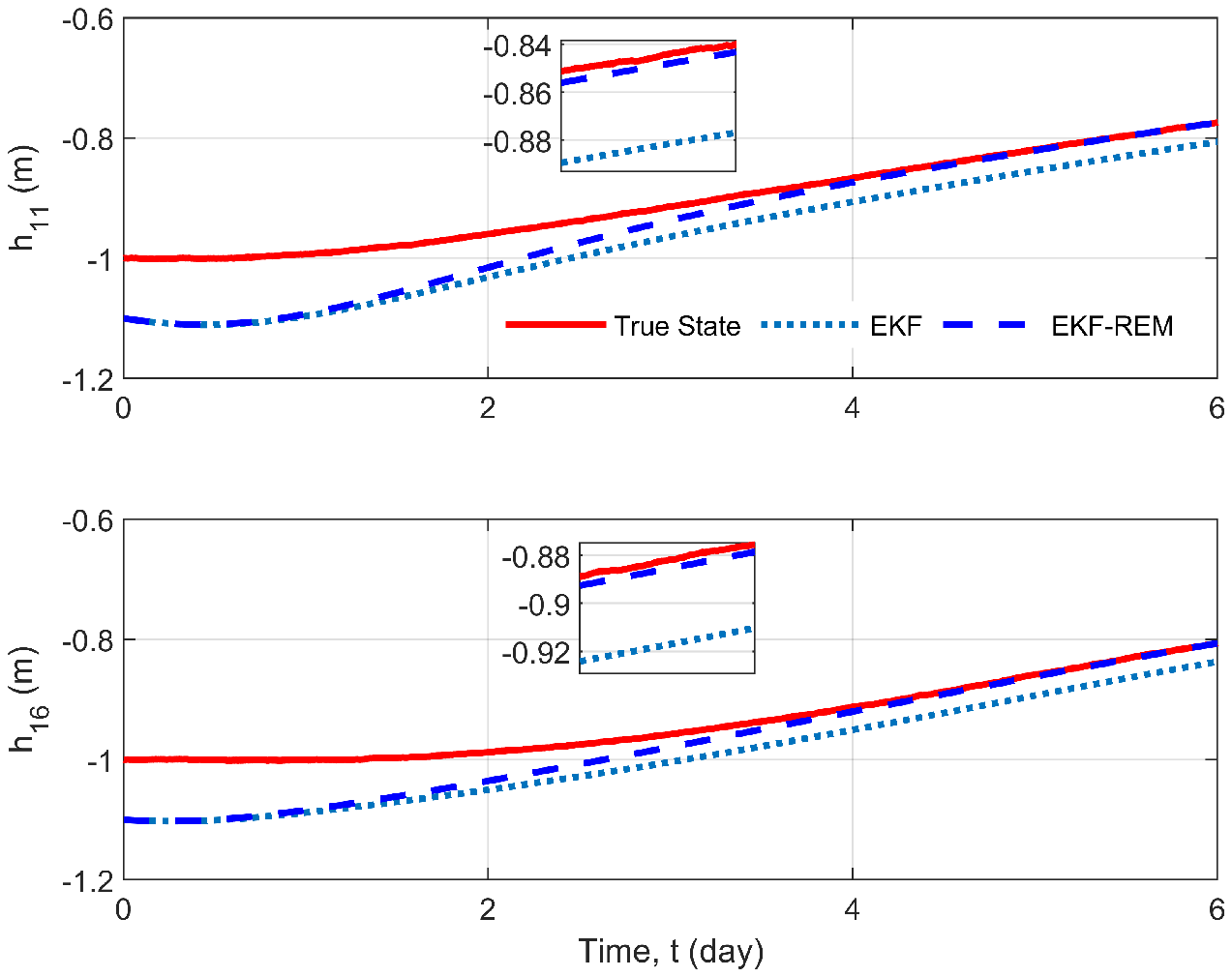}
	\caption{Trajectories of the actual state and the estimates by the standard EKF and the EKF-based REM methods at the depth of $\boldsymbol{20}\ cm \ $ $h_{11}$ and $\boldsymbol{30}\ cm \ $ $h_{16}$.}\label{State_C2}
\end{figure}

\begin{figure}[htbp]
\centering
\subfigure{\includegraphics[width=8cm]{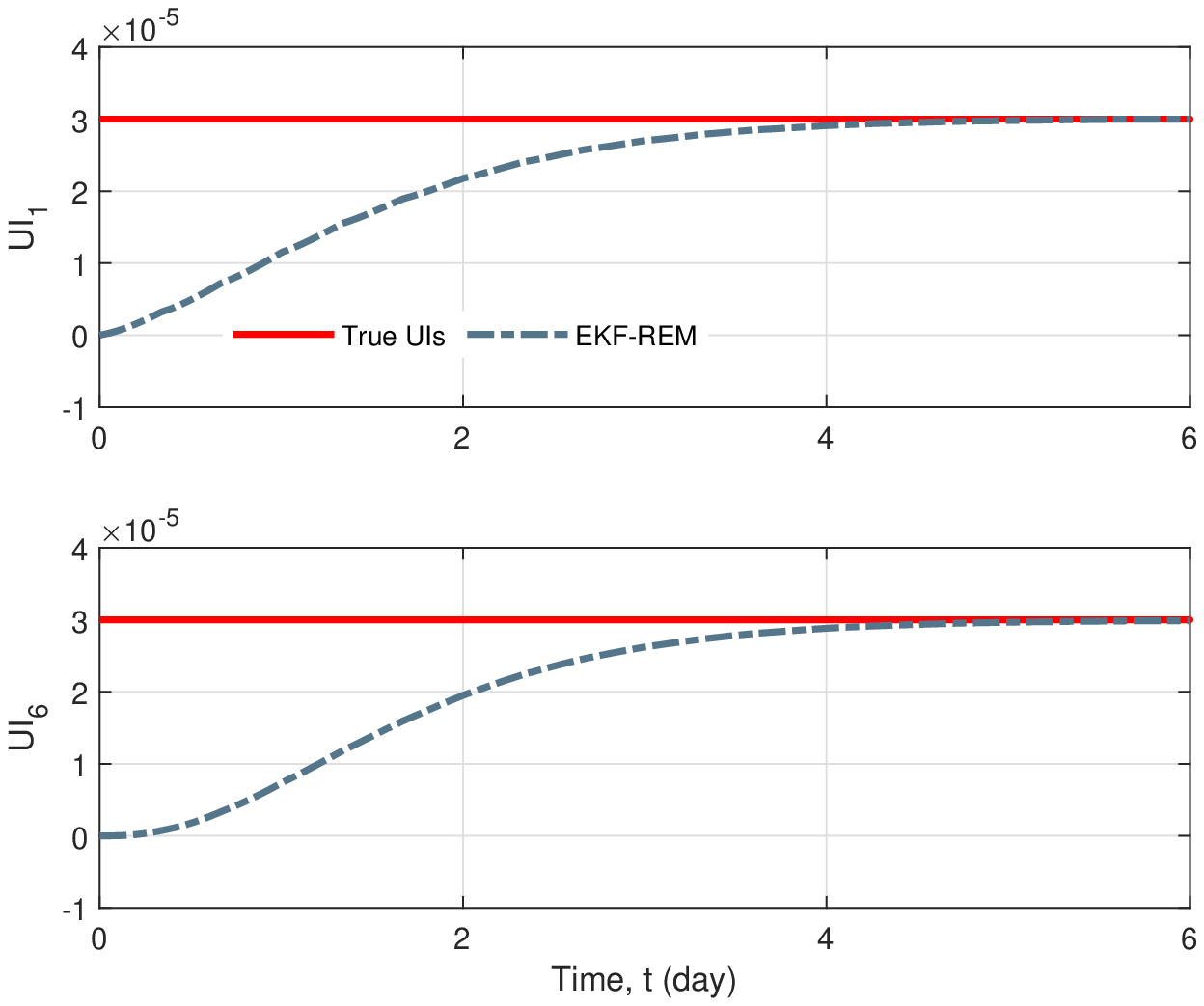}}
\subfigure{\includegraphics[width=8cm]{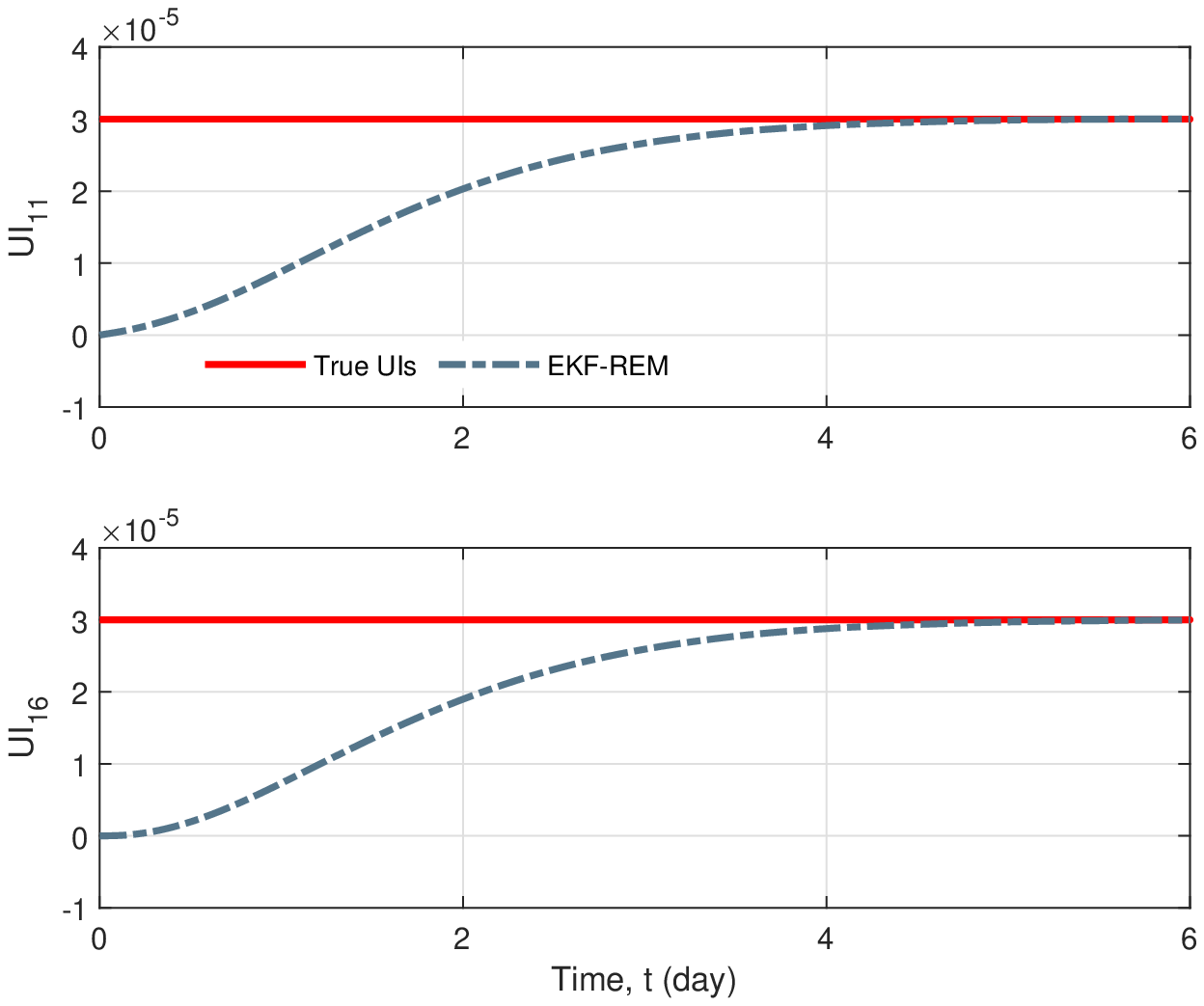}}
\caption{ The actual and the estimated values of the unknown inputs associated with states $h_1$, $h_6$, $h_{11}$, $h_{16}$. }
\label{UIs_C}
\end{figure}

To illustrate the effectiveness of the proposed EKF-based REM algorithm, we present the estimaton for 4 states and compare the performance with a standard EKF without estimating the unknown inputs. Figures \ref{State_C1} and \ref{State_C2} show that the proposed EKF-based REM algorithm converges to the true value after 4 days, indicating that it can provide accurate estimates of the true states. In contrast, the standard EKF approach always exhibits a steady error to the true states. Figure \ref{UIs_C} illustrates that the estimates of the unknown inputs by EKF-based REM algorithm also converge to the true value of $3 \times 10^{-5}$ after about 4 days, indicating that the EKF-based REM algorithm can estimate the unknown inputs correctly. 

\begin{figure}[htbp]
	\centering
	\subfigure{\includegraphics[width=8cm]{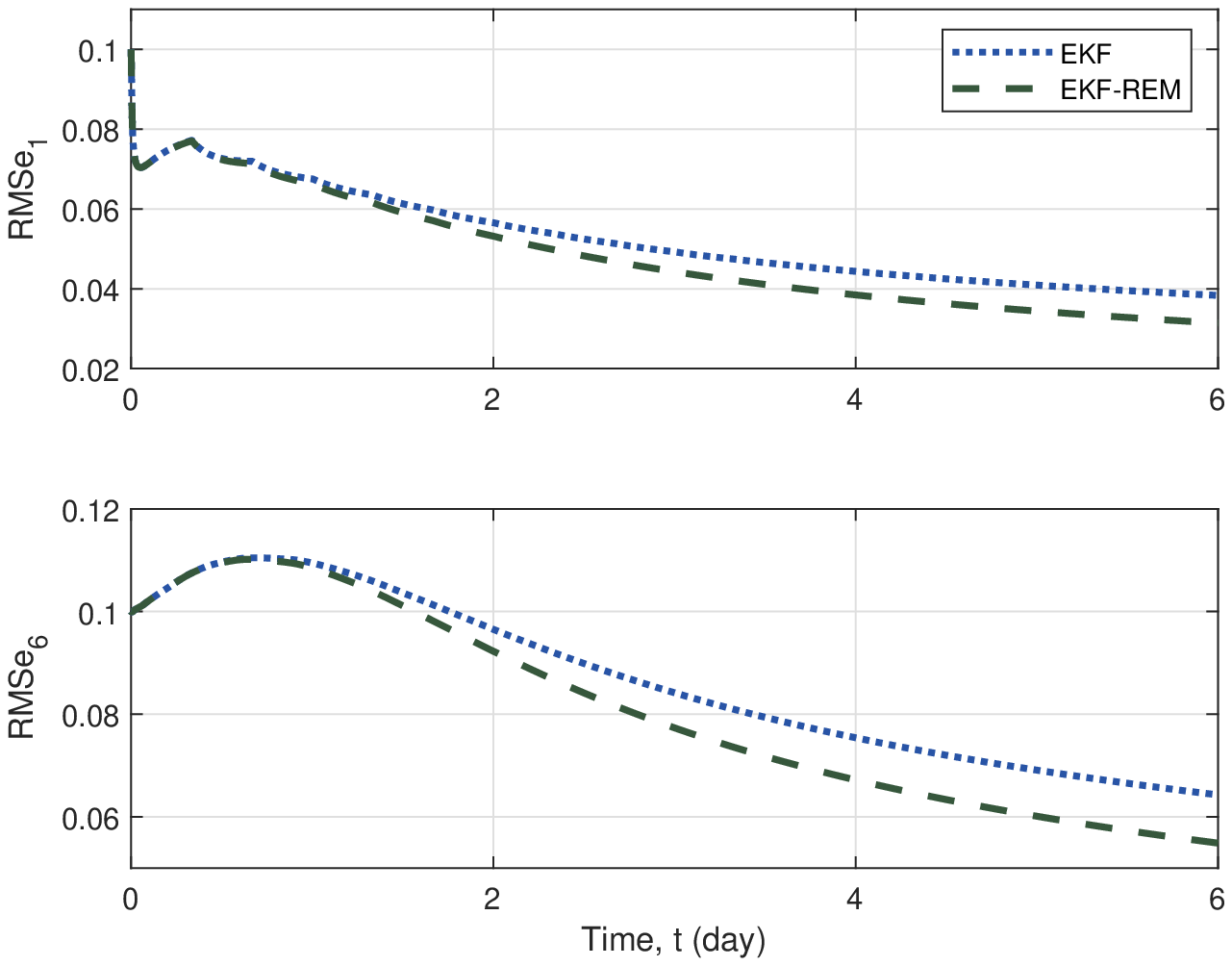}}
	\subfigure{\includegraphics[width=8cm]{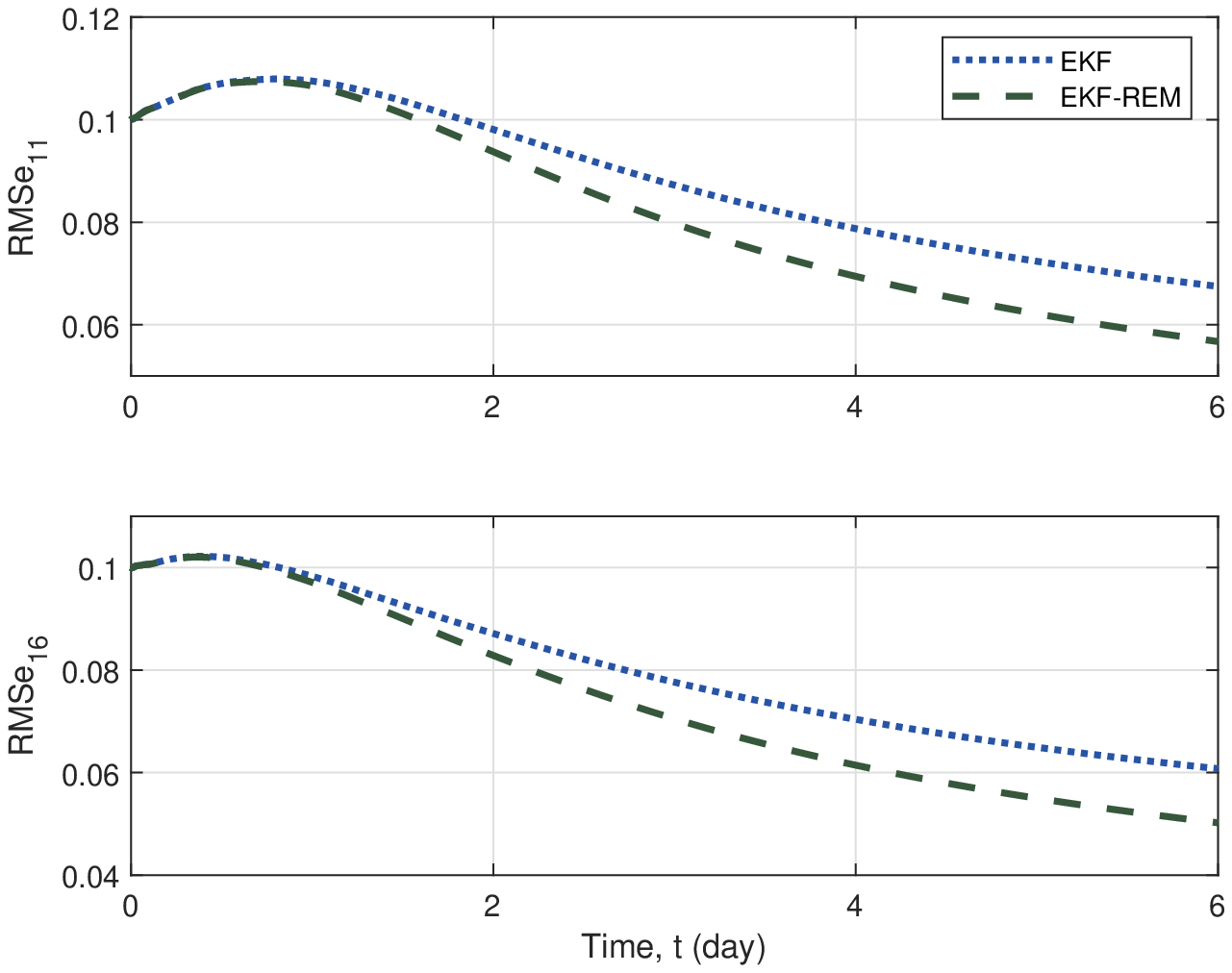}}
	\caption{ Trajectories of RMSEs for state estimation with standard EKF and EKF-based REM algorithm for states $h_1$,$h_6$,$h_{11}$ and $h_{16}$.}
\label{RMSE_C}
\end{figure}

The trajectories of the performance indices RMSE of the EKF and the EKF-based REM are presented in Figure \ref{RMSE_C}, which confirms that the EKF-based REM outperforms the EKF in estimating the states and unknown inputs. These simulations showcase the ability of the EKF-based REM algorithm to handle model mismatch, which appears as a constant bias to the states. Additionally, it should be noted that in this simulation, all the initial guesses of the unknown inputs are consistent. In the subsequent section, we will explore scenarios with different unknown inputs and initial guesses.

\subsection{Scenario 2}
In this scenario, all the true values and initial guesses of the unknown inputs corresponding to each state are different. The true values of unknown inputs for $[h_{1}], [h_6], [h_{11}]$ and $[h_{16}]$ are $2.5 \times 10^{-5}, 3 \times 10^{-5}, 3.5 \times 10^{-5}, 4 \times 10^{-5}$, respectively. The initial guesses for these four unknown inputs are $1 \times 10^{-6}, 6 \times 10^{-6}, 1.1 \times 10^{-5}, 1.6 \times 10^{-5}$, respectively. The true value of the initial state is $x_{0}=-1$, and the initial guess of the state is $\hat{x}_{0}=1.1 \times x_{0}$. The zero mean and variance of $4 \times 10^{-9}$ and $8 \times 10^{-7}$ for process and measurement noise are considered, respectively. 

\begin{figure}[!t]\centering
	\includegraphics[width=11cm]{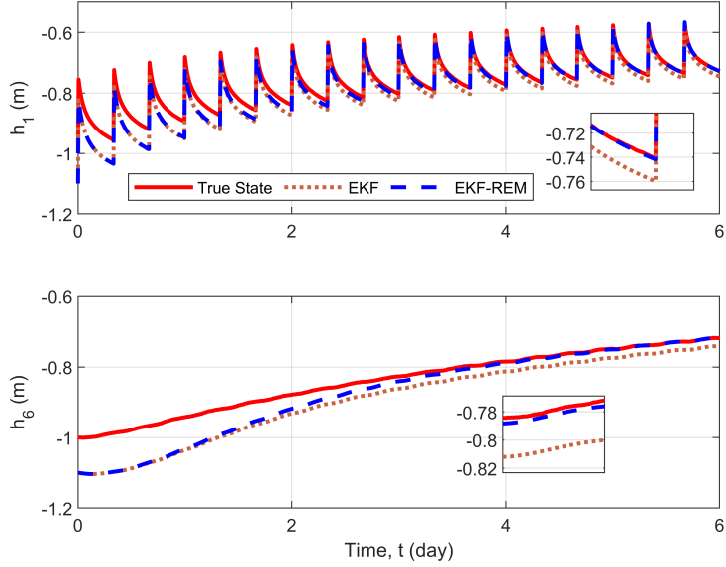}
	\caption{Trajectories of the actual state and the estimates by the standard EKF and the EKF-based REM methods at the surface $h_1$ and depth of $\boldsymbol{10}\ cm \ $ $h_6$.}\label{State_D1}
\end{figure}

\begin{figure}[!t]\centering
	\includegraphics[width=11cm]{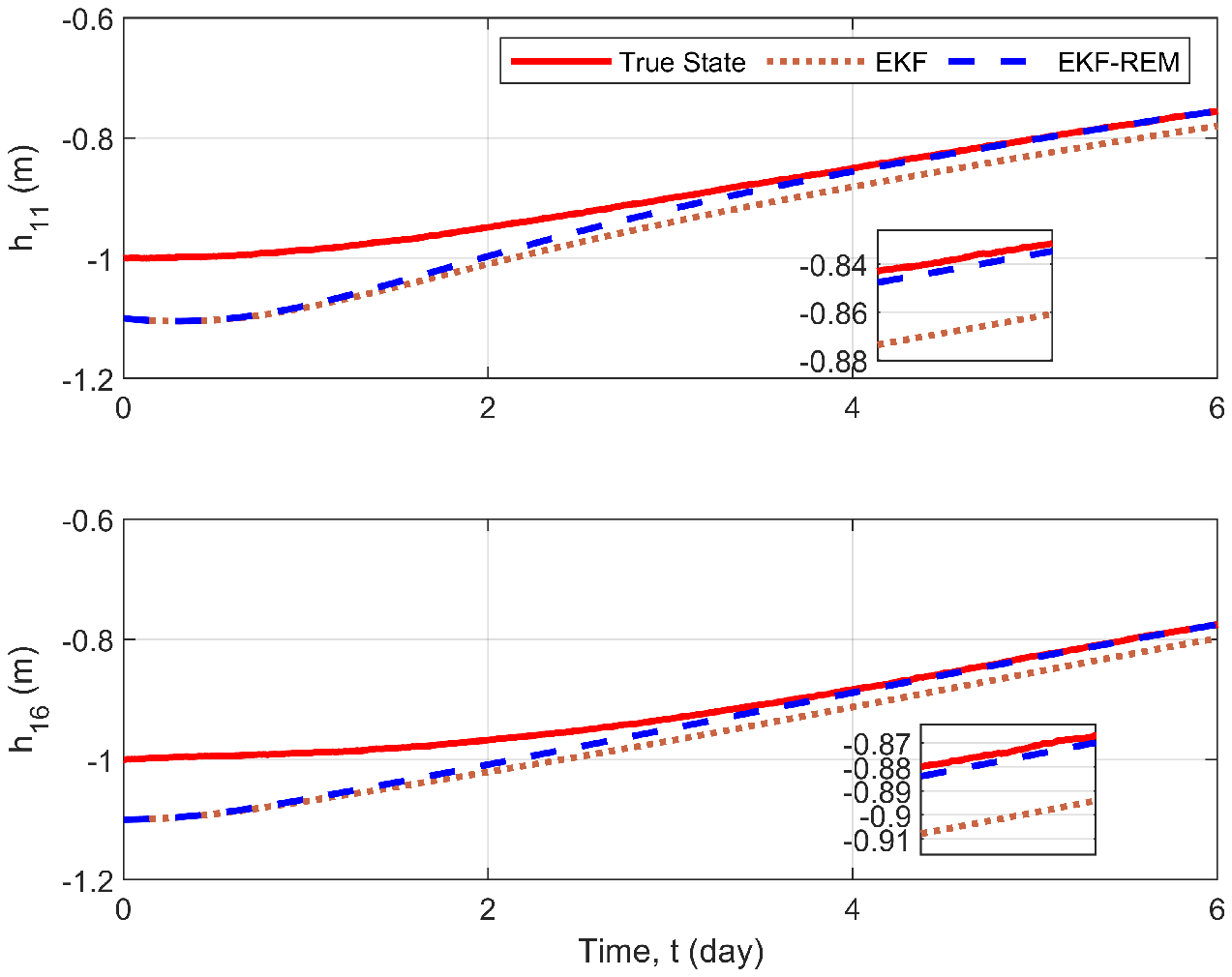}
	\caption{Trajectories of the actual state and the estimates by the standard EKF and the EKF-based REM methods at the depth of $\boldsymbol{20}\ cm \ $ $h_{11}$ and $\boldsymbol{30}\ cm \ $ $h_{16}$.}\label{State_D2}
\end{figure}

\begin{figure}[htbp]
\centering
\subfigure{\includegraphics[width=8cm]{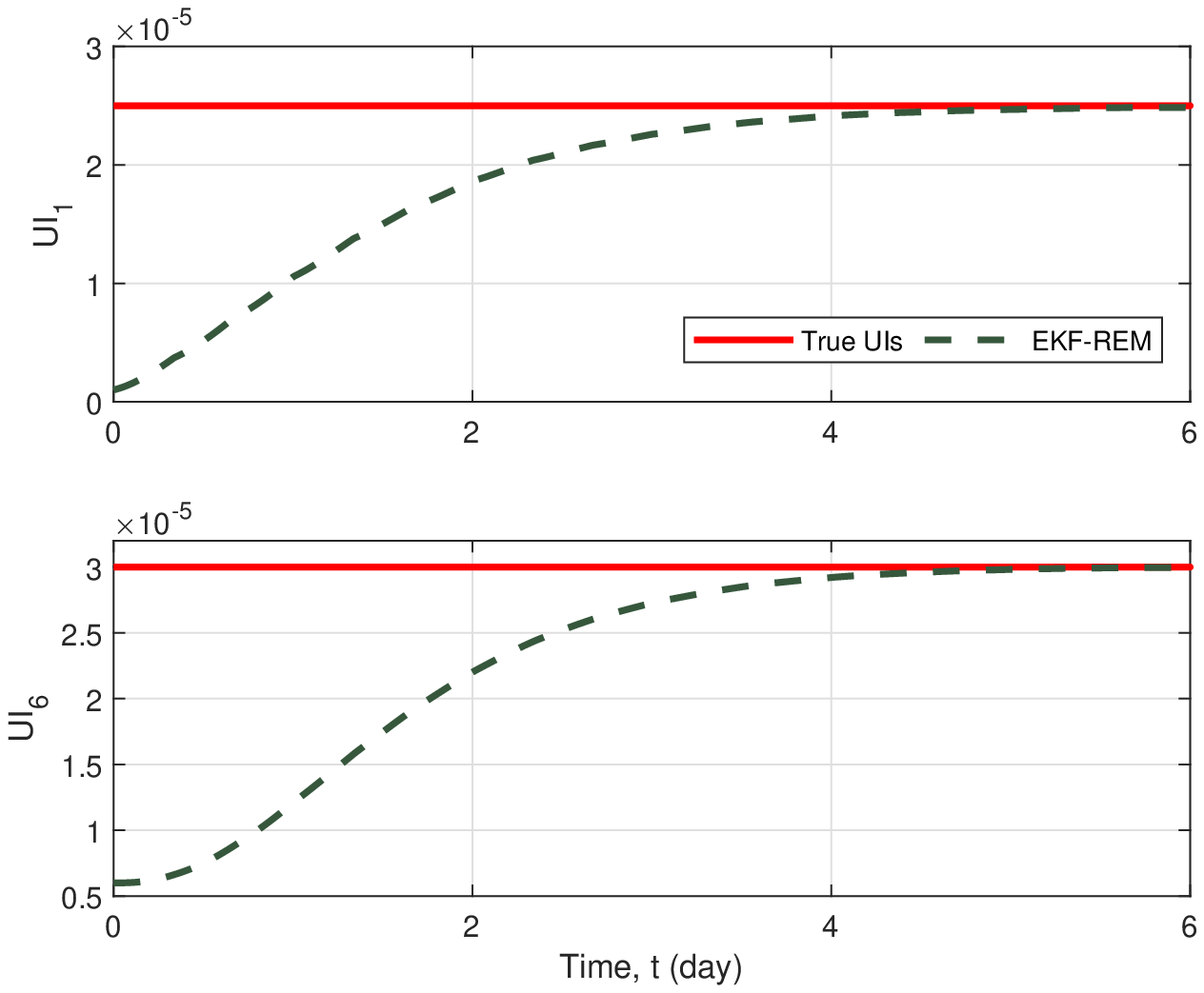}}
\subfigure{\includegraphics[width=8cm]{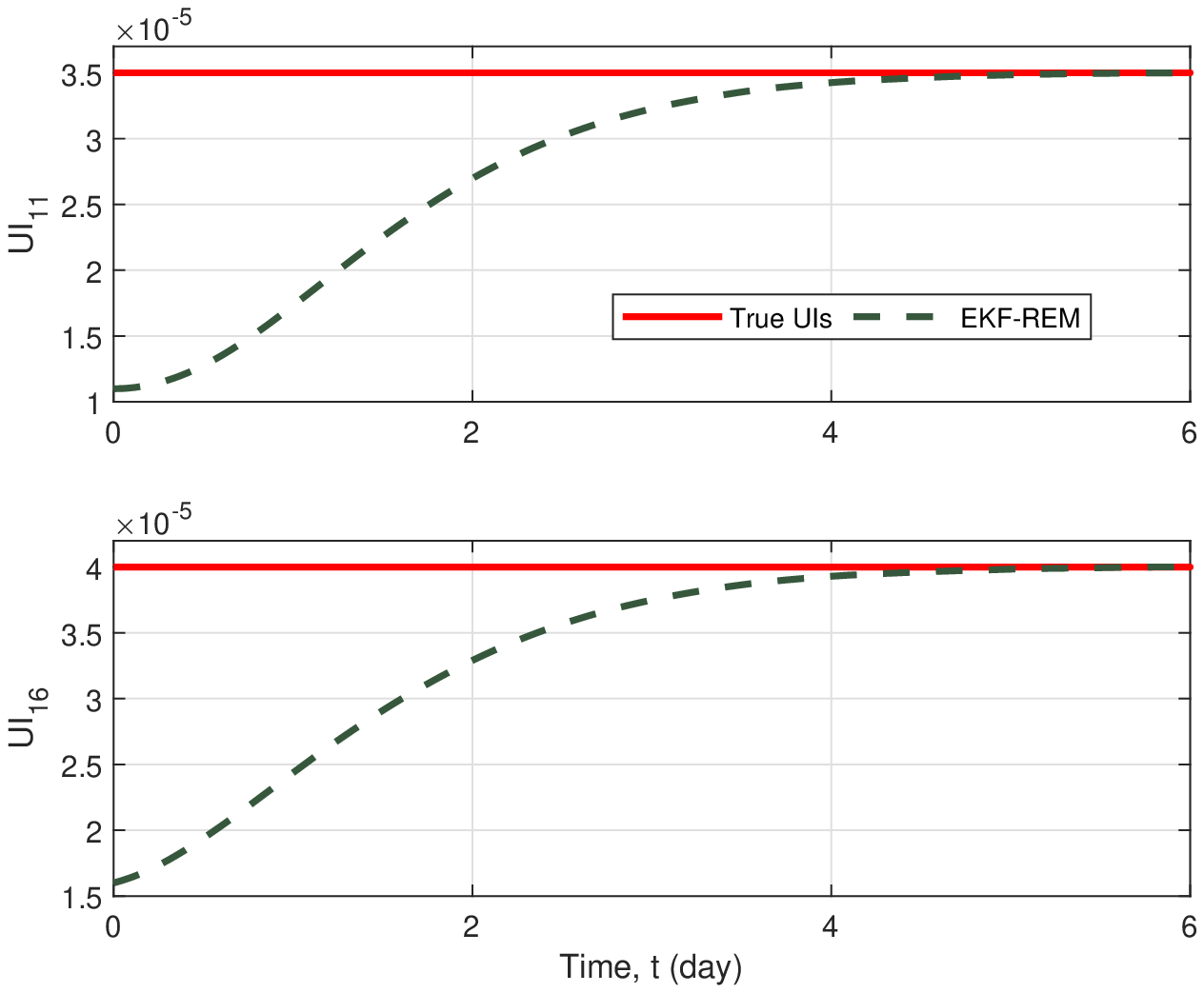}}
\caption{ The actual and the estimated values of the unknown inputs associated with states $h_1$, $h_6$, $h_{11}$, $h_{16}$.}
\label{UIs_D}
\end{figure}

The proposed EKF-based REM algorithm demonstrates its ability to accurately estimate the states in scenarios where the unknown inputs for each state differ, as shown in Figures \ref{State_D1} and \ref{State_D2}. These figures demonstrate that the EKF-based REM algorithm can track the true states after approximately 4 days. On the other hand, the EKF approach performs poorly in such situations, exhibiting a steady error to the true states. Additionally, Figure \ref{UIs_D} shows that the EKF-based REM algorithm can estimate the unknown inputs correctly, even when the true values and initial guesses differ.

\begin{figure}[htbp]
	\centering
	\subfigure{\includegraphics[width=8cm]{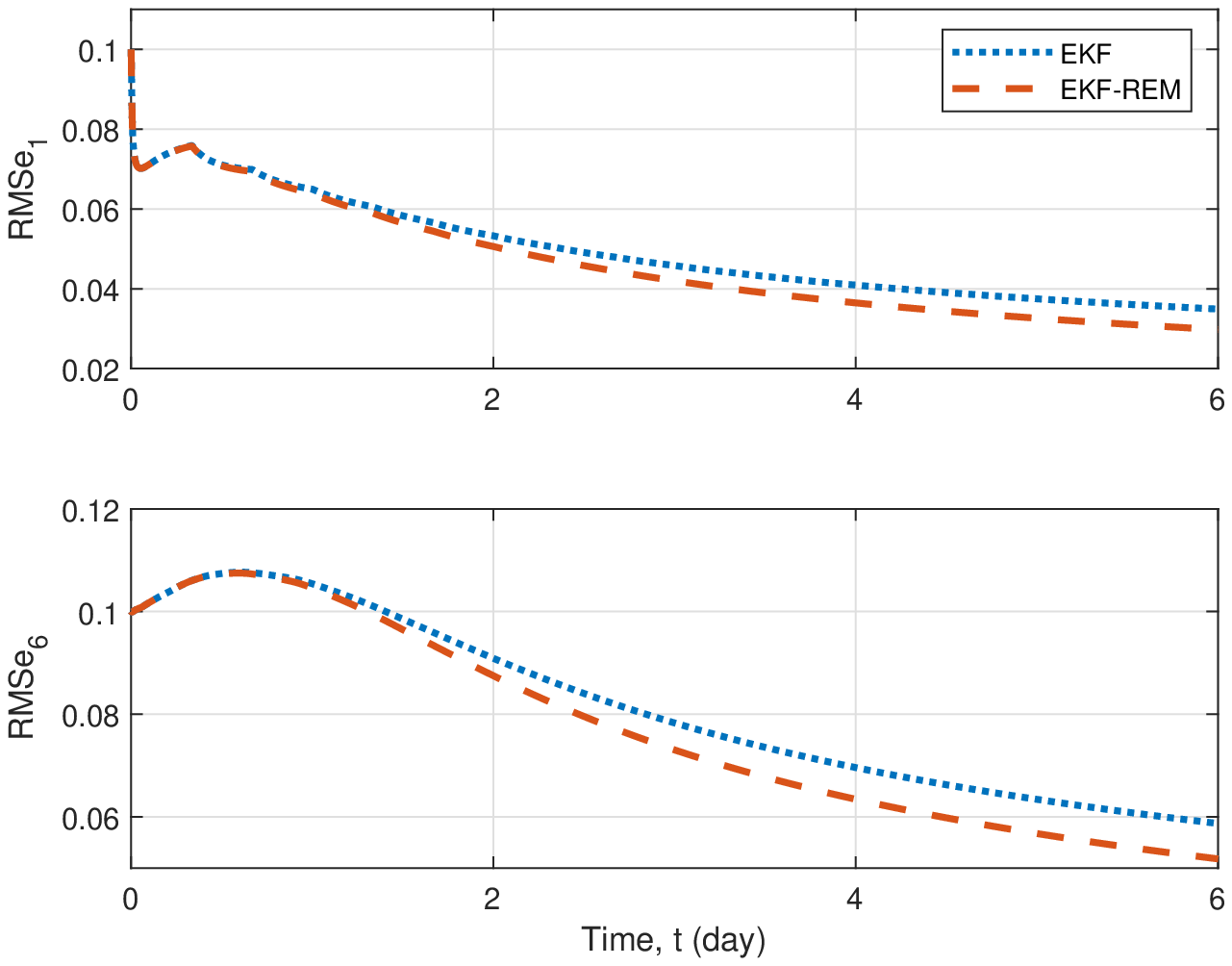}}
	\subfigure{\includegraphics[width=8cm]{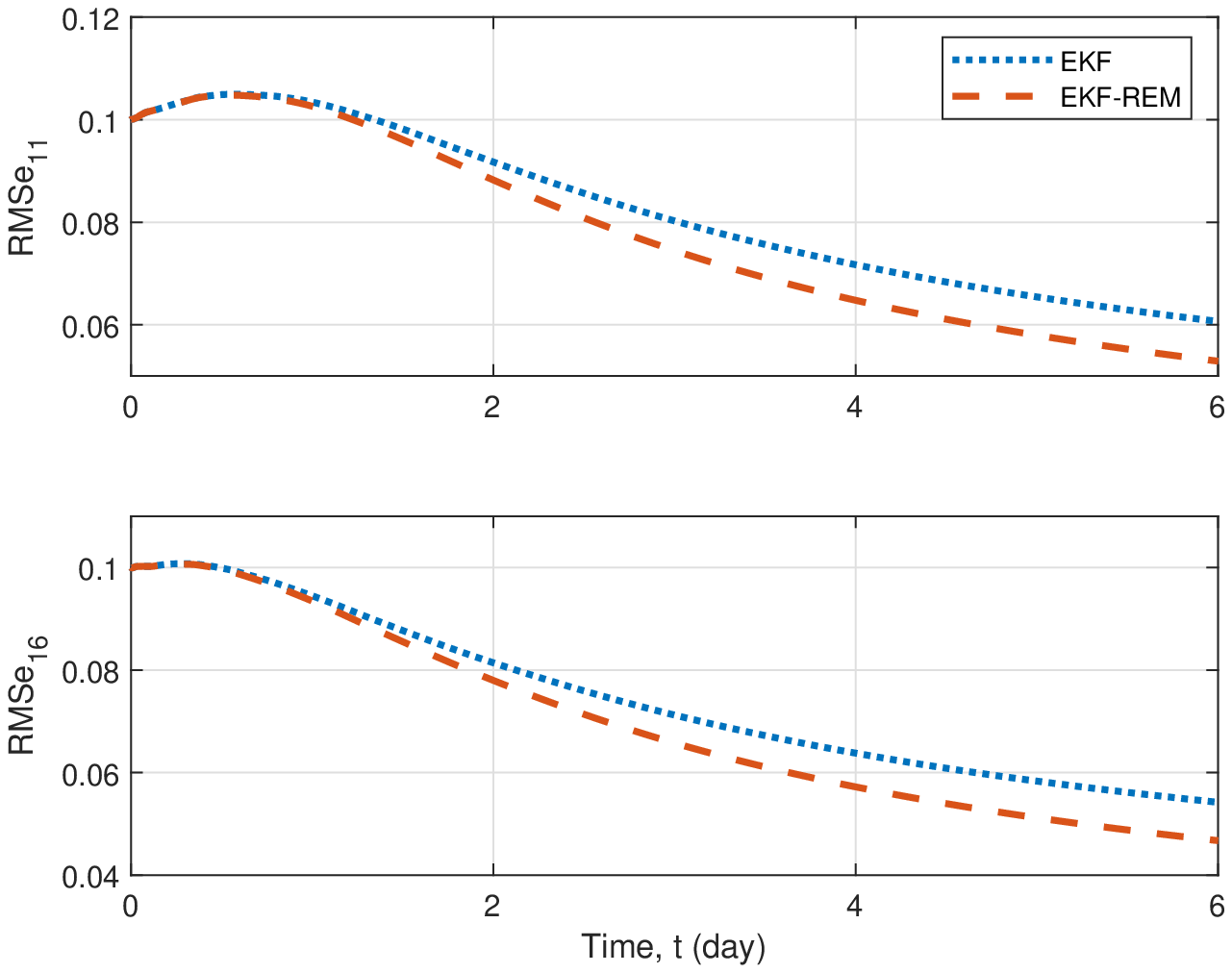}}
	\caption{ Trajectories of RMSEs for state estimation with standard EKF and EKF-based REM algorithm for states $h_1$,$h_6$,$h_{11}$ and $h_{16}$.}
\label{RMSE_D}
\end{figure}

Figure \ref{RMSE_D} shows the trajectories of the RMSE performance indices associated with the EKF and EKF based REM algorithm, which indicate that the EKF based REM algorithm outperforms the EKF when estimating states and unknown inputs. This simulation demonstrates that the EKF-based REM algorithm is capable of handling situations with model mismatches, which result in different constant biases to the states with different initial guesses. In the next section, we will discuss simulations of unknown inputs due to parameter variations.

\subsection{Scenario 3}
In this scenario, the unknown inputs are caused by the parameter variations, which means the reference crop efficiency $Kc$ and evapotranspiration values $Et$ are time-varying and we do not have any prior knowledge of them. The true value of the crop efficiency is $0.88$ from day $1$ to day $3.5$, and after day $3.5$, this value varies to $1.08$. The initial guess of the crop efficient is $1.8$. The reference evapotranspiration value is $1.4 \ mm/day$ from day $1$ to day $3.5$, and $1.5 \ mm/day$ after day $3.5$. The initial guess of the evapotranspiration rate is $1.3 \ mm/day$. The true value of the initial state is $x_{0}=-0.8$, and the initial guess of the state is $\hat{x}_{0}=1.1 \times x_{0}$. The process noise and measurement noise are considered in the simulations and they have zero mean and variance of $4 \times 10^{-9}$ and $8 \times 10^{-7}$, respectively. 

\begin{figure}[!t]\centering
	\includegraphics[width=11cm]{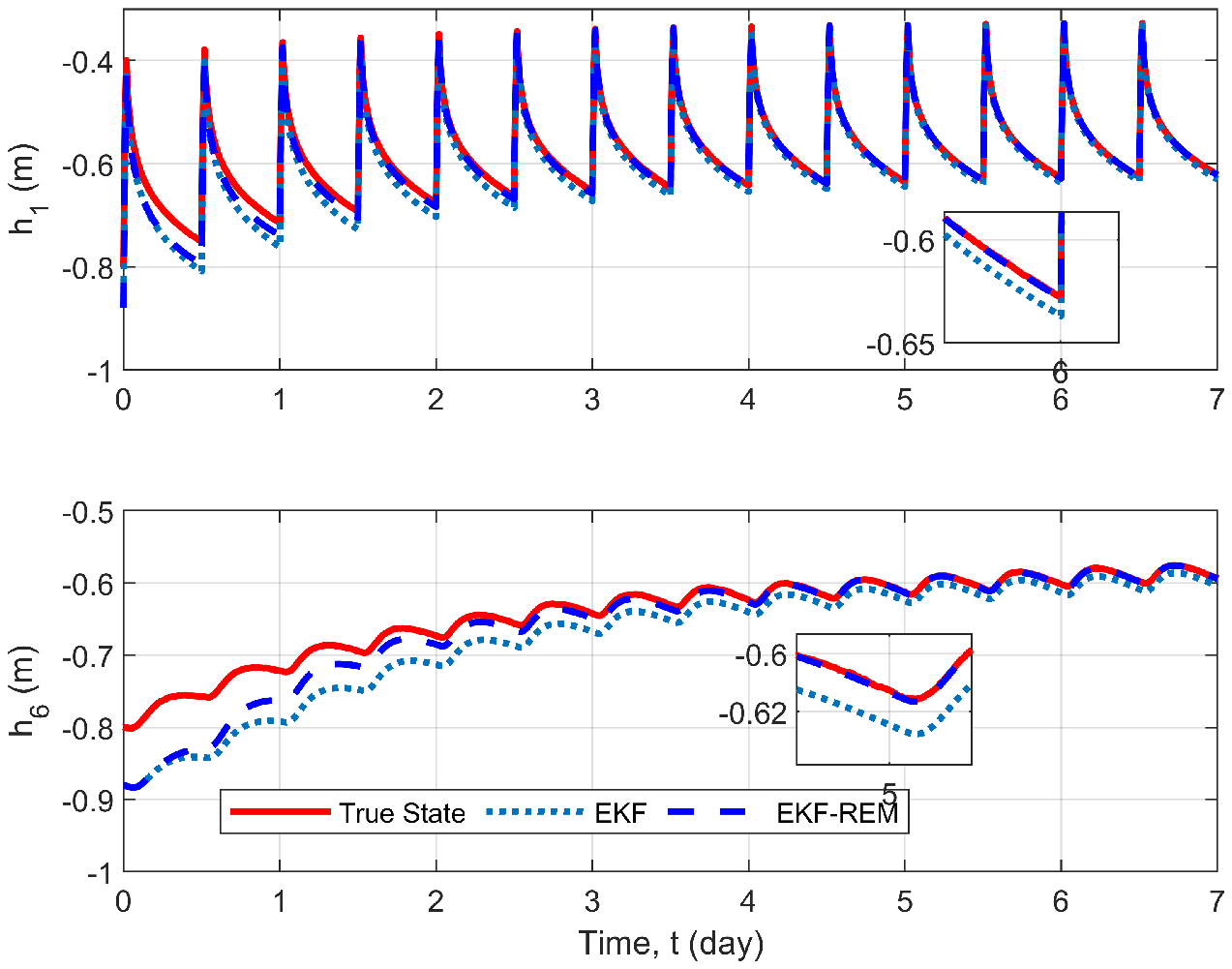}
	\caption{Trajectories of the actual state and the estimates by the standard EKF and the EKF-based REM methods at the surface $h_1$ and depth of $\boldsymbol{10}\ cm \ $ $h_6$.}\label{State_P1}
\end{figure}

\begin{figure}[!t]\centering
	\includegraphics[width=11cm]{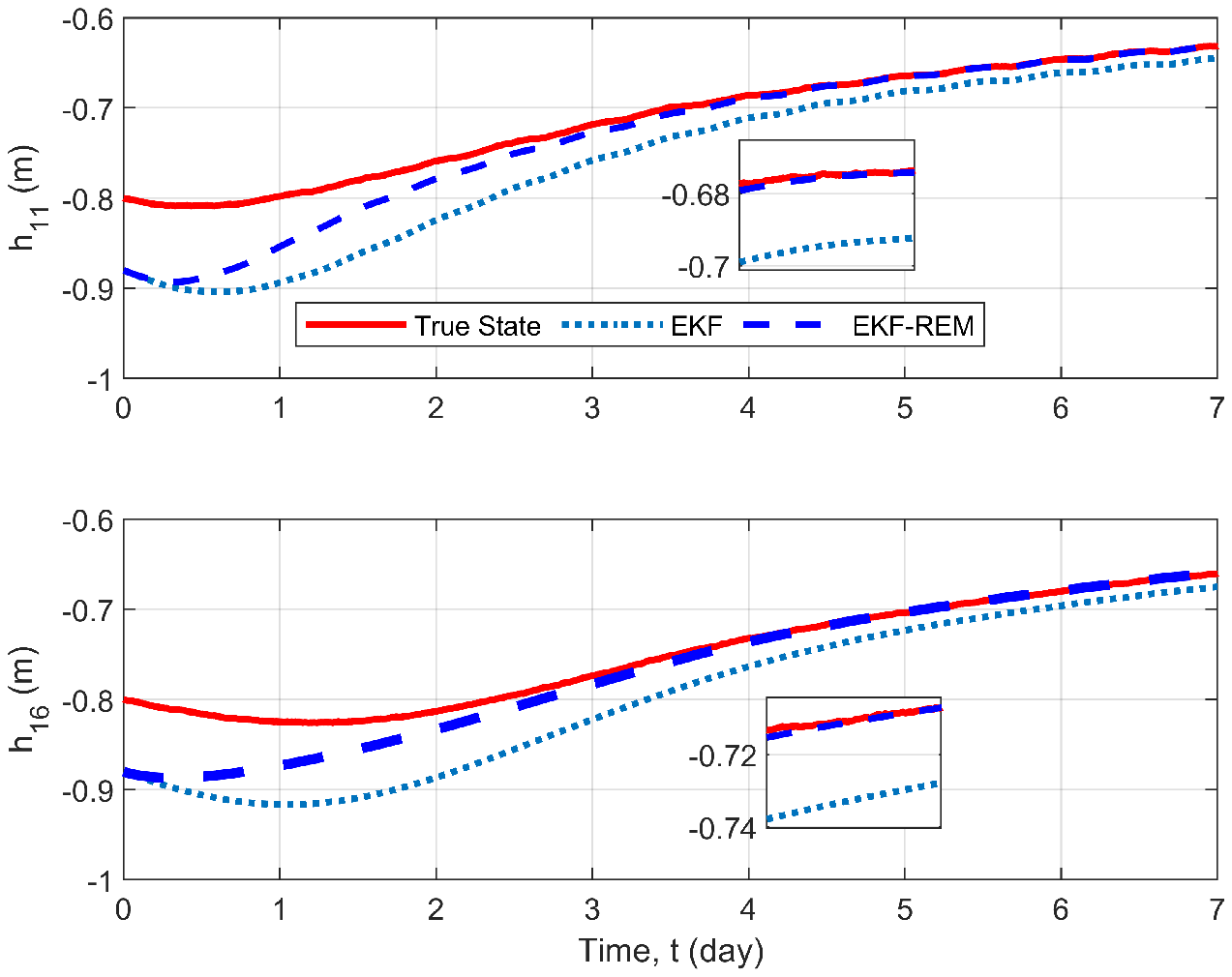}
	\caption{Trajectories of the actual state and the estimates by the standard EKF and the EKF-based REM methods at the depth of $\boldsymbol{20}\ cm \ $ $h_{11}$ and $\boldsymbol{30}\ cm \ $ $h_{16}$.}\label{State_P2}
\end{figure}

Figures \ref{State_P1} and \ref{State_P2}  clearly show that the proposed EKF-based REM algorithm is capable of accurately estimating the true states even when there is parameter variation, within about 3 days. On the other hand, the EKF approach displays poor performance in state estimation and exhibits a steady error when used to estimate states with model mismatch.

\begin{figure}[htbp]
	\centering
	\subfigure{\includegraphics[width=8cm]{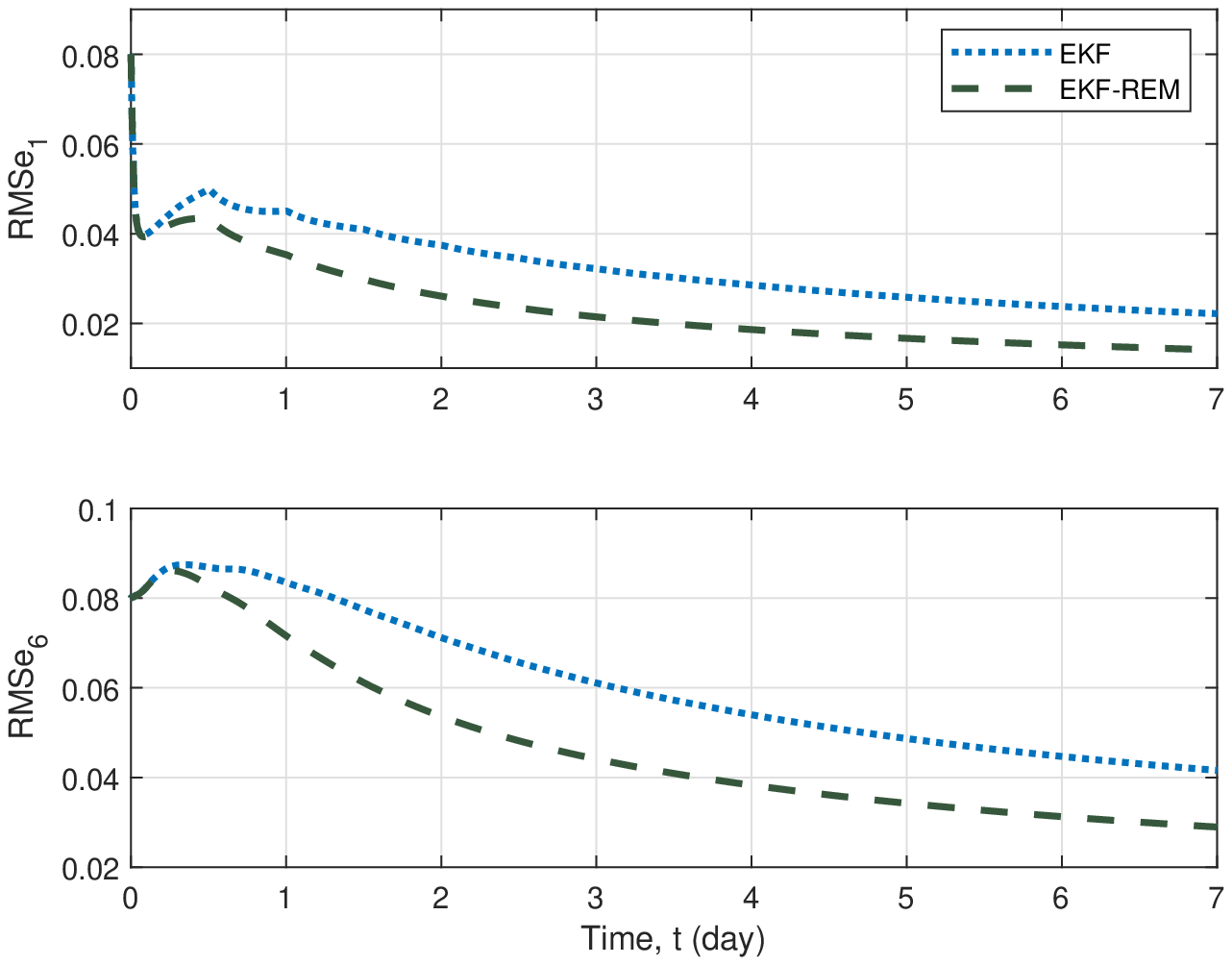}}
	\subfigure{\includegraphics[width=8cm]{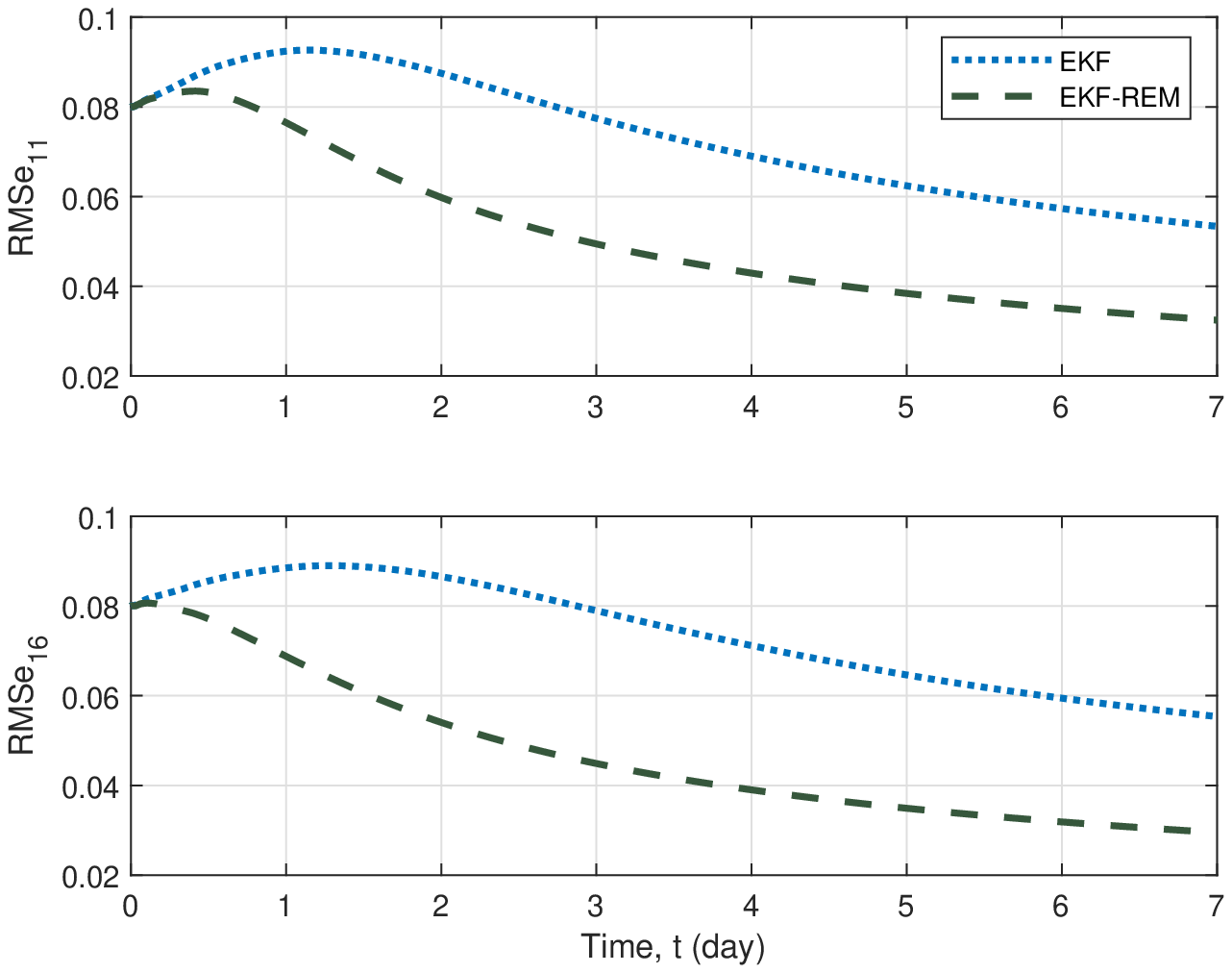}}
	\caption{ Trajectories of RMSEs for state estimation with standard EKF and EKF-based REM algorithm for states $h_1$,$h_6$,$h_{11}$ and $h_{16}$.}
\label{RMSE_P}
\end{figure}

The trajectories of the performance indices RMSE associated with the EKF and EKF-based REM are presented in Figure \ref{RMSE_P}, which confirms that the EKF-based REM algorithm outperforms the EKF model in estimating the states with model mismatch. These results demonstrate that the EKF-based REM algorithm can handle situations where there are errors in the model parameters leading to model mismatch. Moreover, the EKF-based REM algorithm is theoretically capable of addressing model errors resulting from infinite difference approximation.

\section{Conclusion}
This paper addresses the problem of state estimation in the 1D agro-hydrological process with model mismatch. To represent the model mismatch, additive unknown inputs are introduced into the state equation. The minimum number of sensors is determined using sensitivity analysis and orthogonalization methods, which also maximizes the degree of observability. The recursive EM algorithm, derived from the traditional EM algorithm, is used for online state and unknown inputs estimation. In the E-step of the REM algorithm, the EKF is used to update the states and covariance. Simulation results show that the EKF-based REM approach can accurately compensate for model mismatch compared to the EKF algorithm. It is also demonstrated that when model error exists, common filter methods may not have the necessary robustness for accurate state estimation.

\section{Declaration of Competing Interest}

The authors declare that they have no known competing financial interests or personal relationships that could have appeared to influence the work reported in this paper.

\section{Acknowledgment}
This work was supported by the National Natural Science Foundation of China (grant number 61991402, 61973136, 61833007) and China Scholarship Council (CSC).

%\bibliographystyle{unsrt}
%\bibliography{mybibfile}

\end{document}